\documentclass{asi-preprint}
\usepackage{placeins,needspace}
\usepackage{iftex}
\ifPDFTeX
\usepackage[T1]{fontenc}
\usepackage[utf8]{inputenc}
\fi
\usepackage[english]{babel}
\usepackage{amsthm}
\usepackage{asi-fonts}
\usepackage{mathtools}
\usepackage{booktabs,graphicx,enumitem,xcolor}
\usepackage{tikz}
\definecolor{inkblue}{HTML}{154D61}
\definecolor{midgray}{HTML}{46565F}
\usepackage[colorlinks=true,allcolors=inkblue,bookmarksopen=true,pdfdisplaydoctitle=true]{hyperref}
\usepackage{orcidlink}
\providecommand{\doi}[1]{\href{https://doi.org/#1}{doi:\,#1}}
\renewcommand*{\thesection}{\arabic{section}}
\renewcommand*{\thetable}{\arabic{table}}
\renewcommand*{\thefigure}{\arabic{figure}}
\renewcommand*{\thesubsection}{\thesection.\arabic{subsection}}
\makeatletter
\def\p@subsection{}
\renewcommand{\fnum@figure}{\textbf{Figure\ \thefigure}}
\makeatother
\newtheorem{theorem}{Theorem}[section]

\newtheorem{corollary}[theorem]{Corollary}
\theoremstyle{definition}\newtheorem{example}[theorem]{Example}
\theoremstyle{remark}
\newcommand{\paperTitle}{Effective Stabilization of Shortest Addition Chains under Repeated Doubling}
\hypersetup{pdfauthor={Piotr Masierak},pdftitle={Effective Stabilization of Shortest Addition Chains under Repeated Doubling},pdfsubject={Preprint 1.3}}
\begin{document}
\title{\paperTitle}
\author{Piotr Masierak\,\protect\orcidlink{0009-0008-9895-5329}}
\thanks{\href{mailto:pmasierak@patent.pl}{pmasierak@patent.pl}}
\affiliation{Dom Zatorski Scientific Foundation, Chorzów, Poland}
\date[]{November 26, 2026}
\begin{abstract}
Squaring extends an exponentiation plan for $n$ to one for $2n$, but a different plan may avoid the additional multiplication. We give a computable stopping rule for the sequence $\ell(2^tn)-t$, where $\ell$ is the shortest addition-chain length. The proof examines a shortest chain at the first shift meeting a prescribed excess budget. A sufficiently wide gap between run endpoints would allow the same budget to be met one shift earlier. Bounding the width and number of these gaps gives an explicit horizon. Two paths sharing no doubling gate sharpen it to $\lfloor\log_2 n\rfloor+O_a(1)$ for a single target at fixed excess budget $a$. We also give a fixed-parameter algorithm that constructs a compressed chain from the positions of the set bits. Assuming completeness of the published classifications through four small steps, we prove the sharp bound $k\le\lfloor\log_2 m\rfloor-8$ for a drop from more than four to four small steps at shift $k\ge1$ from an odd base $m$, and characterize both equality families. For a fixed set of distinct targets, paths ending at different outputs yield a refined carry bound and an effective joint stabilization horizon. Transposition then extends stabilization to every prescribed positive symbol-count vector and gives an eventually linear penalty for ordering the symbols in separate blocks. The dyadic-ladder case retains a scalar formula.
\end{abstract}
\maketitle

\section{Introduction}
An addition chain describes an exponentiation computation that reuses previously obtained powers. To compute $x^n$ starting with $x$, each multiplication of two available powers adds their exponents. The fewest multiplications in this model is the shortest addition-chain length $\ell(n)$. Squaring $x^n$ gives $x^{2n}$ for one more multiplication. For $n=191$, however, both exponents $191$ and $382$ can be reached optimally in eleven multiplications \cite{knuth,cliftdouble}. Conversely, even if the first hundred doublings of a given $n$ each increase the optimum by one, those observations alone do not justify stopping the search for a later saving.

Altman proved that the sequence $\ell(2^t n)-t$ eventually becomes constant and asked how to decide whether it is constant from the outset \cite[\S1.4]{altman}. An explicit stopping rule must bound the search using the input alone. Section~\ref{dmain:orbit} derives such a rule from the structure of a chain at the \emph{first} saving. If a long run of doublings can be shortened while preserving the saving, the saving would already have been available at the preceding exponent. Bounding the runs that escape this argument gives a numerical horizon. Supplement~S4.1 explains a second, qualitative route from Flammenkamp's polynomial description, following the connection drawn in \cite[\S2.2, Theorem~4]{altmanint}.

The assembly-space formulation of string construction in \cite{rsos} relates minimum assembly cost at a prescribed length to addition chains. The subsequent study of optimal joint assembly spaces \cite{jointspace} distinguishes collective optimization from requiring each target to have an individually optimal plan. We use the same distinction for several scalar targets. This provides the connection between the scalar results developed here and the string-assembly application in Section~\ref{bridge:sec}, whose model and transposition argument are given explicitly.

The algorithm in Section~\ref{sec:algorithms} makes the stopping rule effective even when the exponent is supplied as a short list of occupied bit positions. Section~\ref{chain:paths} gives a second horizon by bounding the doublings used by two dependency paths. The resulting coefficient of the binary length is sharp if the published small-step catalogues are complete. Section~\ref{dmain:level} states that assumption and identifies the equality cases. Section~\ref{bridge:sec} transfers stabilization to string assembly, retaining an exact scalar formula for dyadic ladders and extending the result to arbitrary prescribed counts. Supplement~S7 proves the refined horizon for fixed target sets and the resulting block-order formula. Supplement~S6.6 separates fixed-pattern optimization from independent certificate checking; its exact assignment implementation removes the factorial enumeration in that subproblem.

\paragraph{}
Arithmetic conventions.
An addition chain is $1=a_0<a_1<\cdots<a_r=n$, with $a_i=a_j+a_h$ for some $j,h<i$; the operands may coincide. Its length is $r$, so $\ell(1)=0$. Independent computational steps may be sorted by their positive output values. Write $\lambda(n)=\lfloor\log_2 n\rfloor$, $w(n)$ for the Hamming weight (number of ones) of the binary expansion, and $s(n)=\ell(n)-\lambda(n)$. A step that leaves $\lambda$ unchanged is a \emph{small step}. Each step increases $\lambda$ by at most one, so an optimal chain contains $s(n)$ small steps. This integer is the ceiling of the real defect $\ell(n)-\log_2 n$ studied by Altman \cite{altman}. A \emph{non-doubling} is a step other than twice the immediate predecessor; its count need not equal the number of small steps.

\section{Computation at small excess}\label{sec:algorithms}
A run of a thousand doublings can be described by its initial value and the number $1000$. The difficulty arises when a later sum uses a value from inside the run. For a budget $k$ close to $\lambda(n)=\lfloor\log_2 n\rfloor$, few steps can depart from doubling the preceding term, an observation used in the enumerations of Flammenkamp and Clift \cite{knuth,flammenkamp,cliftsmall}. We record which runs supply the two operands of each such step. The bits of the target restrict where the operands can lie, so we can then optimize all run lengths together.

\begin{theorem}
Decision and construction at bounded excess.\label{thm:alg-excess}
Given $n\ge1$ and $k\ge\lambda(n)$, write $\sigma=k-\lambda(n)$ and $D=1+\lambda(n)$. Whether $\ell(n)\le k$ can be decided, and a witnessing chain constructed if one exists, in
\[
       2^{O((\sigma+1)^2)}\operatorname{poly}\bigl(D+\log(k+2)\bigr)
\]
bit operations. If the input is the list of positions of its $w$ nonzero bits, where $w=w(n)$, then decision and construction of a \emph{compressed} witness take
$2^{O((\sigma+1)^2)}\operatorname{poly}(w\log(D+1)+\log(k+2))$ bit operations.
\end{theorem}

\begin{proof}
For the growth estimate, write $\varphi=(1+\sqrt5)/2$ and start with the pair $(1,0)$. If $(a_i,a_{i-1})\le A(1,\varphi^{-1})$ coordinatewise, an immediate doubling preserves this bound with $2A$ in place of $A$. At any other step the new maximum is at most $a_i+a_{i-1}\le\varphi A$, and the previous maximum is at most $A$. This includes a doubling of an older term. Thus $n\le 2^{r-f}\varphi^f$ for a chain of length $r$ with $f$ non-doublings. Since $\varphi^4<8$, $f\log_2(2/\varphi)\le r-\log_2 n\le\sigma$ gives $f\le4\sigma$.
Contract its doubling runs and distinguish one input of each of the remaining $f$ additions. If $M=2^f$, the output unfolds into at most $M$ powers of two. Each leaf undergoes at most $M-1$ carries on its way to an occupied bit of $n$. Hence all leaf exponents lie in
\[
 C_f(n)=\{b-i:\ b\text{ is a set-bit position of }n,\ 0\le i<M,\ i\le b\}.
\]
Discard $f$ if $w>M$. Otherwise $|C_f(n)|\le M^2$. For each pair of parents, the relative shift of their distinguished paths is a difference of two leaf exponents: extend both paths from that addition to the output along the same suffix.

For each $0\le f\le4\sigma$, enumerate the ordered parent choices and these $f$ shifts. Lemma~S1.1 in the Supplementary Information proves that an integral schedule of cost at most $k$ exists for an enumerated case precisely when its difference-constraint program has optimum at most $k$. Decoding such a schedule gives the witness. There are at most $(f!)^2 M^{4f}$ cases at each $f$. The program uses $O(f+1)$ variables and inequalities. Its coefficients can be formed from at most $M$ exponent positions per gate, with the bit costs proved in Supplement~S1.1. Summing over $f$ gives the two stated bounds.
\end{proof}

The binary method reaches $n$ within $\lambda(n)+w-1$ steps. Testing these budgets in ascending order gives $\ell(n)$ with parameter $w$. Section~\ref{dmain:orbit} applies the same algorithm to $2^tn$. In the sparse input format, multiplication by $2^t$ simply adds $t$ to each occupied bit position. For each excess budget $a$, the finite-horizon test there decides whether any shift meets it. The least successful budget gives the eventual excess, and binary search gives the first shift attaining it (S1.2). A different task calls for all lengths up to $X$. Combining a window construction with the chain count of De Koninck, Doyon and Verreault \cite{dkdv} gives time $X^{1+o(1)}$, measured in the number of entries produced (S1.3).

For a finite set $T$ of positive integers, let $L(T)$ be the fewest additions, starting from $1$, needed to produce every target in $T$. Such a computation is a \emph{joint addition sequence}, also called a joint addition chain. Its increasing list of distinct values is an addition chain for its largest value. Targets may share intermediate sums, and repeated target values occur only once in $T$. In the terminology of \cite{jointspace}, $L(T)$ is the collective optimum for the unary ensemble $\{\sigma^n:n\in T\}$ over any fixed basic symbol $\sigma$: it equals its collective assembly index (CASI). We extend this minimum-size convention to empty and singleton ensembles, with basic targets free, so $L(T)=0$ for $T\subseteq\{1\}$ and $L(\{n\})=\ell(n)$. The ancestor plan of each target need not be individually optimal.

For a prescribed target set, the recurrence on $L(T)$ can also certify a lower bound by excluding all possible final splits; see Supplement~S2. A related decision problem supplies an entire mandatory prefix as part of the input. Supplement~S3 proves NP-completeness for a variable suffix budget $r$ and W[1]-hardness when parameterized by $r$, even when the target lies strictly between the last prefix entry and its double. The exponent of the prefix length in the exact extension algorithm is optimal under the coloured $(r+1)$-SUM hypothesis for each fixed $r\ge2$. The prefix can encode choices which are not compulsory when starting from $1$. Neither that reduction nor the recurrence decides whether unrestricted single-target addition chains are NP-hard.

\section{A finite stopping rule for repeated doubling}\label{dmain:orbit}

To compare different shifts, subtract the $t$ steps that merely double the starting target. This gives $b_n(t)=\ell(2^tn)-t$. Appending one doubling proves $b_n(t+1)\le b_n(t)$. Write $s(n)=\ell(n)-\lambda(n)$ and let $w(n)$ count the ones in $n$. Then $b_n(t)=\lambda(n)+s(2^tn)$. A \emph{small step} leaves the position of the leading bit unchanged. Each other step advances it by one, so a shortest chain has $s(n)$ small steps. A sum such as $5=3+2$ can advance the leading bit too.

The integer $b_n(t)$ is bounded below by $\lambda(n)$ and eventually attains its minimum $\ell_{\rm st}(n)$. Let $K(n)$ be the first shift attaining it. After $K(n)$ one optimal chain suffices for the whole orbit, with one doubling appended per subsequent shift. The number of steps other than immediate doublings stays fixed within this family. This observation motivated the question of how to recognize when all possible savings have occurred. Altman proved eventual stability and asked for an effective test of $K(n)=0$ \cite[\S1.4]{altman}. Supplement~S4.1 derives a qualitative test from Flammenkamp's effective description cited by Altman \cite[\S2.2, Theorem~4]{altmanint}.

To obtain an explicit stopping rule, consider the \emph{first} shift that meets a given excess budget $a$. Even before stabilization, such a chain has at most $4a$ departures from immediate doubling (S1.1). If a stretch of doublings could be shortened while keeping the same excess, the budget would already have been met one shift earlier. We bound how far the run endpoints can be spaced before such a shortening becomes possible.

\begin{theorem}
A finite horizon for the full length.\label{dmain:horizon}
For \(n\ge1\) and integer \(a\ge0\), set
\[
 F=4a,\qquad U=2^F-1,\qquad H(n,a)=(2F+1)(\lambda(n)+U)+U .
\]
There exists \(t\ge0\) with \(s(2^t n)\le a\) if and only if \(s(2^{H(n,a)}n)\le a\).
In particular \(K(n)\le H(n,a_0)\) for any certified bound \(s(n)\le a_0\); one may always take \(a_0=w(n)-1\).
\end{theorem}

\begin{proof}
Choose the first feasible shift $t$. At $t=0$ the first-attainment bound is automatic. For $t>0$ let $m=\lambda(n)$ and take a shortest chain of length at most $m+t+a$ for $N=2^tn$. By the growth estimate in Section~\ref{sec:algorithms} it has $f\le4a$ steps other than immediate doublings. Choose the predecessor doubling whenever such a decomposition exists. Each gate of a shortest chain is used in computing the output.

\paragraph{}
Coordinates for the runs.\\
The \(f\) marked gates delimit \(f+1\) runs of lengths \(h_i\ge0\), \(0\le i\le f\). Write \(v_i\) for the value at the start of run \(i\). For some \(p_j,q_j<j\),
\[
 v_0=1,\quad v_j=2^{u_j}v_{p_j}+2^{z_j}v_{q_j},\qquad
 0\le u_j\le h_{p_j},\quad 0\le z_j\le h_{q_j},\qquad
 N=2^{h_f}v_f .
\]
All gates have a path to the target: otherwise removing unused computations would shorten a shortest chain. Distinguish the first parent at each sum and define
\[
 \rho_0=0,\quad \rho_j=\rho_{p_j}+u_j,\quad E_i=\rho_i+h_i,\quad
 d_j=\rho_{q_j}+z_j-\rho_j,\quad c_i=2^{-\rho_i}v_i .
\]
Then \(c_0=1,\ c_j=c_{p_j}+2^{d_j}c_{q_j}\) and \(N=2^{E_f}c_f\). For the fixed ordered parents and offsets \(d_j\), an integer endpoint assignment is feasible if
\begin{equation}
 \rho_0=0,\quad 0\le\rho_i\le E_i,\qquad
 \rho_{p_j}\le\rho_j\le E_{p_j},\quad
 \rho_{q_j}\le\rho_j+d_j\le E_{q_j}.             \label{dsi:endpoints}
\end{equation}
Indeed, reconstruct the two shift counts as \(\rho_j-\rho_{p_j}\) and \(\rho_j+d_j-\rho_{q_j}\). Both are integers in their parent runs. Induction on \(j\) gives a program on positive integers with output \(2^{E_f}c_f\) and cost \(f+\sum_i(E_i-\rho_i)\). To make a conventional increasing chain, keep only ancestors of the output, merge equal values, then sort distinct values in increasing order. Each noninitial value has a decomposition into two strictly smaller retained values. This never increases length.

\paragraph{}
A bound on offsets from binary carries.\\
Unfold the \(f\) sums of the run diagram into a tree. There are at most \(M=2^f\) leaves, each contributing a nonnegative integer power of two to \(N\). Merge two equal powers to obtain one power with exponent one greater; continue until the ordinary binary expansion remains. At most \(M-1\) merges occur. Every final bit has an exponent in \([t,t+m]\), so every original exponent lies in \([t-(M-1),t+m]\). Following only distinguished edges gives leaf exponent \(E_f\). For any fixed gate \(j\) choose a path from it to the output; follow that same suffix after using the all-distinguished path to either of its parents. The two full path exponents differ by \(d_j\). In particular
\[
   |d_j|\le C:=m+M-1,\qquad E_f-t\ge-(M-1).                \tag{H1}
\]
The choice of a path upwards is legitimate because each gate is used. The nonnegativity assertion concerns the original leaf exponents, not the normalized offsets, which can be negative.

\paragraph{}
Gap lemma.\\
Let \(t\ge1\) and suppose two consecutive endpoint coordinates in \([0,E_f]\) are \(L<B\), with \(B-L>C\). Decrease every endpoint at least \(B\) by one, including those beyond \(E_f\). The anchor remains zero. Every constraint in~\eqref{dsi:endpoints} has the form \(x-y\le c\) with \(|c|\le C\), using the anchor for the lower bound zero. The only adverse change is \(x\le L\), \(y\ge B\). In that case \(x-(y-1)\le L-B+1\le-C\le c\). Thus all operand constraints persist.

No run length increases. Along the distinguished path to the output, consecutive intervals meet, their union contains \([0,E_f]\), and no endpoint lies strictly between \(L\) and \(B\). Hence at least one interval crosses the gap and loses a doubling. The coefficients \(c_i\) are unchanged, so the new output is \(2^{E_f-1}c_f=2^{t-1}n\). Sorting the resulting program gives a chain with at least one fewer addition. When \(E_f>(2f+1)C\), such a gap exists by counting the at most \(2f+2\) endpoints between 0 and \(E_f\). This also covers \(f=0\) and \(C=0\).

This contradicts first attainment. Hence $t=0$ or
\[
 t\le(2f+1)(m+2^f-1)+2^f-1\le H(n,a).
\]
Appending $H(n,a)-t$ doublings now reaches the test shift. Conversely, success at that shift supplies a witness. Apply the bound to the eventual excess $a=\ell_{\rm st}(n)-m\le a_0$ and use monotonicity of $H$ in $a$. The binary method gives $s(n)\le w(n)-1$.
\end{proof}

\begin{corollary}
A finite test for stability.\label{dmain:stabilitytest}
Let \(H=H(n,w(n)-1)\). Then
\[
 \bigl[\ell(2^t n)=\ell(n)+t\ \text{for every integer }t\ge0\bigr]
 \quad\Longleftrightarrow\quad \ell(2^H n)=\ell(n)+H .
\]
\end{corollary}
\begin{proof}
The theorem gives \(K(n)\le H\). Thus \(b_n(H)\) is the eventual minimum of the nonincreasing sequence \(b_n\); it equals its initial value precisely when the sequence is constant throughout.
\end{proof}

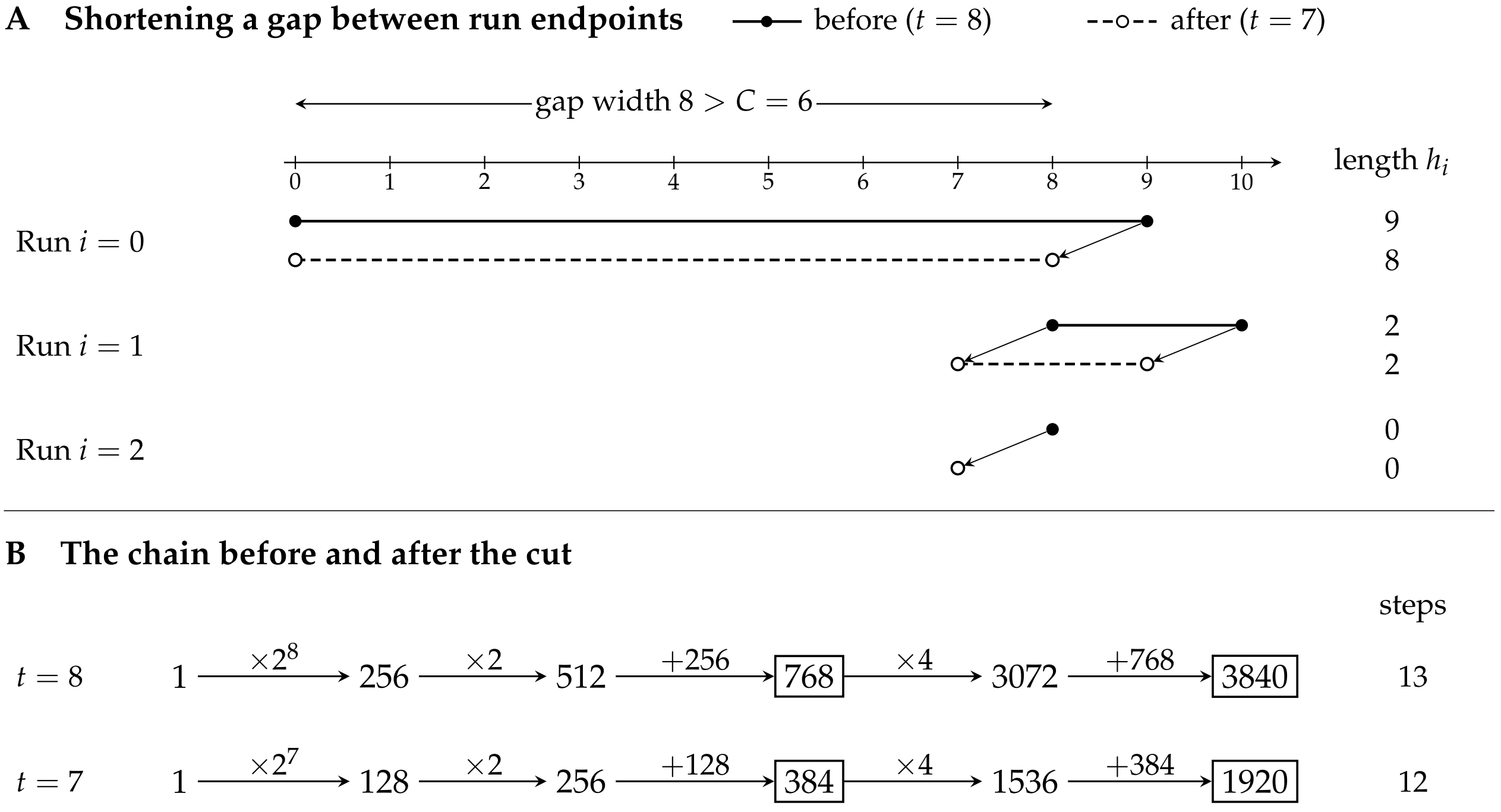
\begin{figure}[htbp]
\centering
\resizebox{.98\linewidth}{!}{%
\begin{tikzpicture}[
  x=1cm,y=1cm,>=stealth,
  title/.style={anchor=west,font=\normalsize\bfseries,inner sep=0pt},
  rowname/.style={anchor=west,font=\small,inner sep=0pt},
  key/.style={font=\small,inner sep=1pt},
  old/.style={line width=.8pt},
  new/.style={line width=.8pt,dash pattern=on 3pt off 1.6pt},
  move/.style={->,line width=.34pt,shorten <=2pt,shorten >=2pt},
  state/.style={font=\normalsize,inner xsep=2.7pt,inner ysep=2.6pt},
  sum/.style={state,draw,line width=.65pt},
  op/.style={midway,above=1pt,font=\small,inner sep=0pt},
  next/.style={->,line width=.55pt}
]
\path[use as bounding box] (0,-6.95) rectangle (16.1,1.64);
\node[title] at (.12,1.46) {A\quad Shortening a gap between run endpoints};
\draw[old] (7.82,1.46)--(8.55,1.46);
\fill (8.18,1.46) circle (1.7pt);
\node[rowname] at (8.68,1.46) {before ($t=8$)};
\draw[new] (11.57,1.46)--(12.30,1.46);
\draw[fill=white,line width=.55pt] (11.94,1.46) circle (1.7pt);
\node[rowname] at (12.44,1.46) {after ($t=7$)};
\draw[<->,line width=.45pt] (3.2,.59)--node[midway,fill=white,key]{gap width $8>C=6$} (11.2,.59);
\draw[->,line width=.45pt] (3.08,-.03)--(13.62,-.03);
\foreach \j in {0,...,10}{
  \draw[line width=.4pt] ({3.2+\j},-.10)--({3.2+\j},.04);
  \node[font=\scriptsize,below=1pt,inner sep=0pt] at ({3.2+\j},-.10) {$\j$};
}
\node[key] at (14.79,-.03) {length $h_i$};
\foreach \y/\i/\lold/\rold/\lnew/\rnew in
  {-.65/0/0/9/0/8,-1.75/1/8/10/7/9,-2.85/2/8/8/7/7}{
  \pgfmathsetmacro{\yn}{\y-.41}
  \node[rowname] at (.24,{\y-.21}) {Run $i=\i$};
  \draw[old] ({3.2+\lold},\y)--({3.2+\rold},\y);
  \draw[new] ({3.2+\lnew},\yn)--({3.2+\rnew},\yn);
  \ifnum\rold>\rnew
    \draw[move] ({3.2+\rold},\y)--({3.2+\rnew},\yn);
  \fi
  \ifnum\lold<\rold\relax
    \ifnum\lold>\lnew\relax
      \draw[move] ({3.2+\lold},\y)--({3.2+\lnew},\yn);
    \fi
  \fi
  \foreach \j in {\lold,\rold}{\fill ({3.2+\j},\y) circle (1.85pt);}
  \foreach \j in {\lnew,\rnew}{\draw[fill=white,line width=.65pt] ({3.2+\j},\yn) circle (1.85pt);}
  \node[key] at (14.79,\y) {$\the\numexpr\rold-\lold\relax$};
  \node[key] at (14.79,\yn) {$\the\numexpr\rnew-\lnew\relax$};
}
\draw[line width=.27pt] (.12,-3.71)--(15.87,-3.71);
\node[title] at (.12,-4.15) {B\quad The chain before and after the cut};
\node[key] at (15.01,-4.75) {steps};
\foreach \line/\yy/\t/\pow/\p/\twop/\v/\fourv/\target/\cost in
  {one/-5.46/8/8/256/512/768/3072/3840/13,two/-6.57/7/7/128/256/384/1536/1920/12}{
  \node[rowname] at (.24,\yy) {$t=\t$};
  \node[state] (base\line) at (1.98,\yy) {$1$};
  \node[state] (p\line) at (4.14,\yy) {$\p$};
  \node[state] (d\line) at (6.22,\yy) {$\twop$};
  \node[sum] (v\line) at (8.63,\yy) {$\v$};
  \node[state] (fv\line) at (10.91,\yy) {$\fourv$};
  \node[sum] (w\line) at (13.34,\yy) {$\target$};
  \draw[next] (base\line)--node[op] {$\times 2^{\pow}$} (p\line);
  \draw[next] (p\line)--node[op] {$\times 2$} (d\line);
  \draw[next] (d\line)--node[op] {$+\p$} (v\line);
  \draw[next] (v\line)--node[op] {$\times 4$} (fv\line);
  \draw[next] (fv\line)--node[op] {$+\v$} (w\line);
  \node[key] at (15.01,\yy) {$\cost$};
}
\end{tikzpicture}
}%
\caption{Shortening a gap in a chain for $15\cdot 2^t$. \textbf{A.} At coordinate $x$, the values in runs $0,1,2$ are $2^x$, $3\cdot2^x$ and $15\cdot2^x$, respectively. Diagonal arrows trace the endpoints that move. The first run loses a doubling, the second shifts as a whole, and the last remains a point. \textbf{B.} Boxes mark the results of the two additions other than doubling the preceding term. An arrow labelled $\times 2^j$ denotes $j$ successive doublings, so $\times4$ counts as two steps.}\label{fig:internal-gap}
\end{figure}

\noindent\textbf{} 
An internal cut.\\
To obtain $15\cdot2^t$, form $v=2^t+2^{t+1}$ and then $v+4v$. Distinguishing the first operand at each sum gives offsets $1$ and $2$. In Figure~\ref{fig:internal-gap}, each interval $[\rho_i,E_i]$ represents a run of $h_i=E_i-\rho_i$ doublings, and a point represents a run of length zero. At $t=8$ the distinct endpoint coordinates are $0,8,9,10$. The gap $(0,8)$ contains no endpoint, although the first run spans the entire gap. Since $8>C=\lambda(15)+2^2-1=6$, every endpoint at coordinate $8$ or higher can move one unit to the left.

\begin{figure}[htbp]
\centering
\begin{tikzpicture}[x=1.27cm,y=0.41cm,font=\small]
 \draw[->,color=midgray] (-.45,0)--(5.6,0) node[right]{$t$};
 \draw[->,color=midgray] (-.45,0)--(-.45,10) node[above,font=\scriptsize]{number of operations};
 \foreach \i/\t in {0/6,1/7,2/8,3/9,4/10,5/11} {
   \draw[gray] (\i,0)--(\i,-.18) node[below,color=black]{\t}; }
 \foreach \y/\v in {1/28,3/30,5/32,7/34,9/36}{
   \draw[gray!25] (-.45,\y)--(5.06,\y);
   \node[left,font=\scriptsize] at (-.47,\y){\v};}
 \draw[very thick,color=inkblue] (0,5)--(1,6)--(2,7)--(3,7)--(4,8)--(5,9);
 \draw[thick,dashed,color=midgray] (0,1)--(1,2)--(2,3)--(3,3)--(4,4)--(5,5);
 \foreach \x/\y in {0/1,1/2,2/3,3/3,4/4,5/5}{
   \fill[midgray] (\x,\y) circle (1.5pt);
   \fill[inkblue] (\x,{\y+4}) circle (1.8pt);}
 \node[anchor=west,font=\footnotesize,color=inkblue] at (5.15,9){$M(n,2n,4n)$};
 \node[anchor=west,font=\footnotesize,color=midgray] at (5.15,5){$\ell(n)$};
\end{tikzpicture}
\caption{Optimal assembly costs at $n=2^t\cdot170055$ under the hypothesis of Theorem~\ref{dmain:sharp}. Each marked point corresponds to one integer shift. Assembling a string with the three prescribed symbol counts costs four more operations than the scalar chain; the two values at $t=8,9$ coincide on each curve.}\label{bridge:figure}
\end{figure}
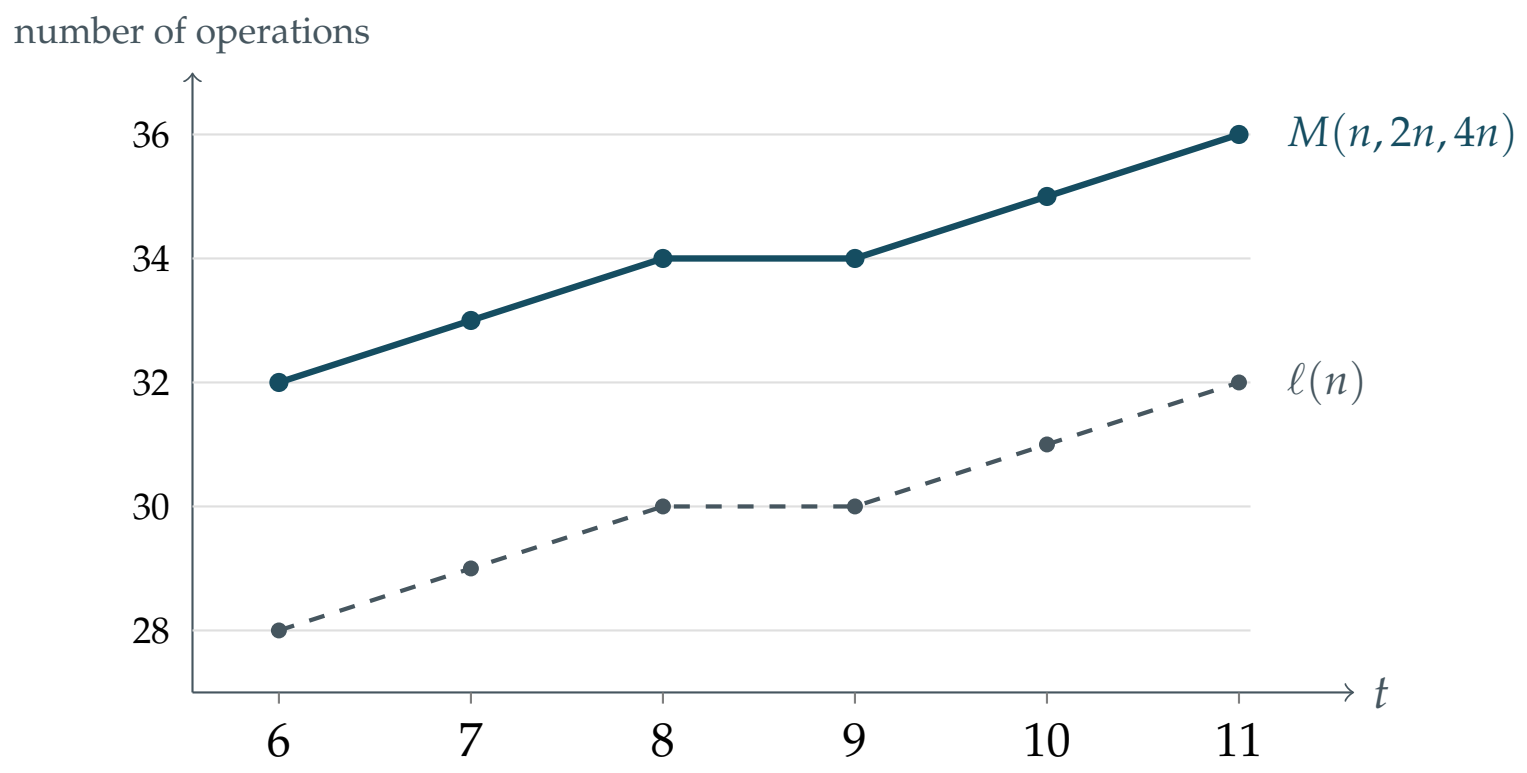

Both rows end with the sum $v+4v$. Shortening the first run thus gives a plan for half the original target in $12$ steps instead of $13$. After subtracting the shift, the two plans have the same cost, $13-8=12-7=5$. If $t$ were the first shift to meet the prescribed excess, cutting the gap would give a plan meeting the same excess already at $t-1$. Figure~\ref{bridge:figure} illustrates an actual drop in $b_n(t)$ on moving to a \emph{larger} target, at the ninth doubling for $n=170055$.

The preceding bound counts the gaps separately. A sharper estimate uses the way a doubling can serve several later sums. If every path from $1$ to the output passes through the same doubling, removing that gate halves the output at the cost of one fewer step. At a first positive shift this is impossible. The obstruction persists when the target consists of $q$ counts chosen separately at each shift.

\section{Two dependency paths and binary carries}\label{chain:paths}

A plan may have several outputs. For $1\le q\le N$ define $F_q(N)$ as $q-1$ plus the shortest length of an addition sequence from which one can select $q$ terms summing to $N$, with repetition. The term $q-1$ counts binary additions needed to sum the selected outputs. Sorting and merging the complete plan proves $F_q(N)\ge\ell(N)$, with equality for $q=1$. To double the sum, start with $1,2$ and execute the same sequence with each term doubled. Thus $F_q(2N)\le F_q(N)+1$. Put $b_{n,q}(t)=F_q(2^tn)-t$ for $q\le n$. This nonincreasing integer sequence is bounded below by $m=\lambda(n)$. Its first minimizing index is denoted by $K_q^{\rm str}(n)$, since the selected terms can be interpreted as symbol counts in a string (Section~\ref{bridge:sec}). Set $v=\nu_2(n)$, the exponent of two in $n$, and $J_0=0$, $J_1=1$, $J_{j+2}=J_{j+1}+J_j$. Use $w(0)=0$.

\Needspace{7\baselineskip}
\begin{theorem}
\label{as:paths10}
Suppose $t>0$ is the first shift satisfying $b_{n,q}(t)\le m+a'$ for integers $0\le a'\le a$. In a shortest addition sequence supplying the $q$ counts at shift $t$, let $f$ be the number of non-doublings and put $c=m+a-(q-1)-f$. Define
\[
 P_c(n)=\begin{cases}\lfloor n/2^c\rfloor+w(n\bmod 2^c),&c\ge0,\\ n2^{-c},&c<0.\end{cases}
\]
Writing $L$ for the number of powers of two in the expanded sum of the $q$ outputs, one has
\begin{align}
 t&\le c-2\min(v,c)+2\bigl(L-P_c(n)\bigr),\label{eq:cap10}\\
 P_c(n)&\le L\le q2^f,\qquad n2^{f+q}\le q2^{m+a}J_{f+3}.\label{eq:fib10}
\end{align}
Maximizing the right side of \eqref{eq:cap10} with $L=q2^f$ over the finitely many $f$ satisfying \eqref{eq:fib10}, including zero as a candidate for the horizon, gives a computable bound $m+O_{q,a}(1)$.
\end{theorem}
\begin{proof}
Represent immediate doublings as unary gates. Expanding the binary sums gives $L$ powers of two, each at most $2^{t+c}$ because the chain contains at most $t+c$ unary gates. Merge equal powers up to that denomination. The final sum for $2^tn$ has $P_c(n)$ terms and smallest exponent $t+\min(v,c)$. Each initial term participates in at most $L-P_c(n)$ merges, so every dependency path contains at least $t+\min(v,c)-(L-P_c(n))$ unary gates.

Join the selected outputs to one sink. A unary gate on every path could be replaced by its input, halving every output and saving at least one step. This contradicts first attainment. Give each unary gate capacity one and all other gates and wires capacity two. A cut of capacity one would consist of a common unary gate. Integral max-flow/min-cut \cite{fordflow} therefore gives two paths sharing no unary gate. Each uses at least $t+\min(v,c)-(L-P_c(n))$ doublings, while the two together use at most $t+c$. Thus $t$ enters twice in the required number of doublings and only once in the available budget. This gives \eqref{eq:cap10}. The largest output is at least $2^tn/q$. A chain of length $R$ with $f$ non-doublings reaches at most $2^{R-f-1}J_{f+3}$. Indeed, a maximal block of $b\ge1$ non-doublings starts after two terms $x/2,x$. Each new term is at most the sum of its two immediate predecessors. Relative to $b$ doublings, the resulting growth factor is at most $J_{b+3}/2^{b+1}$. The identity $2J_{b+c+3}-J_{b+3}J_{c+3}=J_bJ_c$ follows by the Fibonacci recurrence. Merging the bounds for adjacent blocks thus gives the claimed factor for all $f$ non-doublings together. Substituting $R\le t+m+a-(q-1)$ proves \eqref{eq:fib10}. Since $J_{f+3}/2^f\to0$ and $n\ge2^m$, $q,a$ alone bound $f$.
\end{proof}

For a single target, $f\le4a$ and $L\le2^f$ already give the simple bound $t\le m+a+2^{4a+1}$ (S6). Both horizons are linear in $m$ at fixed budget. The coefficient of $m$ thus falls from $8a+1$ to one. If $F_q(n)>m$, set $a=F_q(n)-m-1$ to cover first entry into every strictly lower level, including the final one. If $F_q(n)=m$, the orbit is already stable. The additive constant in Theorem~\ref{as:paths10} can vary with the budget when $a$ is chosen separately for each $n$.

For $191$ the stopping bound and the stopping time can both be checked. Here $q=1$, $m=7$ and $\ell(191)=11$. With the initial excess $a=4$, we have $F=16$, $U=65535$ and $H=33(7+65535)+65535=2\,228\,421$. To find a saving it suffices to test $a=3$, which gives $H=106\,645$. Primality reduces the two-path bound to $119$, the cap on the largest binary power reduces it to $81$, and enumerating gate diagrams reduces it to $20$ (S6). Among the remaining operand choices only $t=1$ survives. There is also a short direct certificate. The number $191$ has seven ones, as does every $191\cdot2^t$. The classical small-step classification gives $s(x)\ge3$ for $w(x)\ge5$, and $s(x)\ge4$ for $w(x)\ge9$ \cite{knuth,altman}. Thus $b_{191}(t)\ge7+3=10$ for all $t$. Since $b_{191}(0)=11$ and $b_{191}(1)=\ell(382)-1=10$, the first shift already attains the permanent lower bound. Thus $K(191)=1$.

A shift-independent lower bound can certify an index without enumerating diagrams. If $B(x)\le\ell(x)$ and $B(2x)=B(x)+1$, appending a chain for a multiplier $h$ scaled by $2^tn$ gives $b_n(t)\ge B(hn)-\ell(h)$. For the weight bound of S6, the resulting maximum over all multipliers is $B_{\times}(n)$. For example, $3\cdot12623=37869$ has top bit $15$ and ten ones, giving $b_{12623}(t)\ge15+4-2=17$. The lengths at $t=0,1,2,3$ are $18,19,20,20$. The normalized cost first meets the bound at $t=3$, so $K(12623)=3$.

Multiplier lower bounds certify $4\,716$ scalar indices among the $8\,192$ odd $n<2^{14}$. Gate enumeration adds $268$, giving $4\,984$ determinations independent of the S5 completeness assumption. For all odd $n<1024$, only $191$, $701$ and $743$ have $K=1$, and the other $509$ have $K=0$. Conditional on catalogue completeness, a separate census using S5 gives counts $(8\,119,69,2,2)$ at indices $0,1,2,3$, respectively, throughout the larger range. Each of the four cases with $K\ge2$ also has a certificate independent of S5. The finite multiplier optimization, certificates, and two methods for redistributing doublings on fixed diagrams are given in S6.

\paragraph{}
Prescribed outputs and separate carry budgets.\\
When the values of several outputs are fixed, each output has its own expansion into powers of two. Two paths ending at different prescribed outputs therefore charge two different carry budgets, whose sum is bounded by the total budget once. The two paths need not be disjoint on non-doubling gates. This refines the direct use of one common carry bound for both paths and gives an effective stopping rule for every prescribed target set.

\begin{theorem}
Effective stabilization of prescribed targets.
\label{v13:joint-stability}
Let $T\subset\mathbb Z_{>0}$ be finite and nonempty. Put
\[
 b_T(t)=L(2^tT)-t,\qquad
 L_{\rm st}(T)=\min_{t\ge0}b_T(t),\qquad
 K_T=\min\{t:b_T(t)=L_{\rm st}(T)\}.
\]
Then $K_T$ is finite and effectively computable, and
\[
 L(2^tT)=t+L_{\rm st}(T)\qquad(t\ge K_T).
\]
For $h=|T|\ge2$ and $m=\lambda(\max T)$, let $H_T(a)$ be the explicit finite maximum in Supplement~S7.1. Any integer upper bound $A\ge L(T)$ certified by a joint addition sequence gives $K_T=0$ if $A=m$, and
\[
 K_T\le H_T(A-m-1)\qquad\text{if }A>m.
\]
For a singleton, the scalar horizon of Theorem~\ref{dmain:horizon} applies.
\end{theorem}

The full proof is in Supplement~S7.1. Its first-attainment argument gives two paths with distinct prescribed endpoints by assigning capacity one to their terminal arcs. Expanding the outputs separately counts each selected output's carry loss at most once. The Fibonacci bound restricts the number of non-doublings to finitely many cases. To pass to later shifts, prepend doublings and scale the entire program, preserving every required output. The target set is fixed throughout; the selected counts in $F_q(2^tn)$ may instead vary with $t$.

\section{Results at three and four small steps}\label{dmain:level}
The published small-step classifications of Knuth, Flammenkamp and Clift \cite{knuth,flammenkamp,clift,cliftsmall,cliftfour} allow us to test the sharpness of the coefficient one in the two-path bound. Their completeness is assumed throughout this section. A row of these catalogues describes a family of binary bit positions by arithmetic conditions. Doubling adds one to every position. A first entry must therefore cross the boundary of a row, even when its parameters are arbitrarily large. Supplement~S5.2 provides two independent checks of that boundary calculation. The public companion includes the covering rows and verification programs.

At level three, for every \(n\) and every \(t\ge1\),
\[
 \min\{4,s(2^t n)\}=\min\{4,s(2n)\}.
\]
Consequently, if \(s(2n)\le4\), then \(2n\) is stable. The five-bit consequences are particularly explicit. Among all numbers of binary weight five, the unstable ones are exactly
\[
 n=2^a+2^{a-7}+13,\quad a\ge13;\qquad
 \ell(n)=a+4,\quad \ell(2^tn)=a+t+3\quad(t\ge1).
\]
A second criterion involves no doubling test. If the five occupied positions of \(n\) are congruent modulo some \(d\ge3\), then \(s(n)=3\) precisely when four distinct positions admit a partition into two pairs of equal sum; otherwise \(s(n)=4\). For instance, the supports \(t+d\{0,1,2,3,8\}\) and \(t+d\{0,1,2,4,8\}\) have the same top bit and weight, yet their costs differ by one at every \(t\ge0\).

For level four set \(c_m(t)=\min\{5,s(2^tm)\}\) and let \(K_4(m)\) be the first index from which \(c_m\) is constant. A displayed value \(5\) in this profile means \emph{at least five}. We use odd bases \(m\); this includes every orbit up to a shift of origin.

\begin{theorem}
The last possible entry into level four.
\label{dmain:sharp}
Assume that for each \(j=0,1,2,3,4\) the published small-step classifications describe the integers \(N\) with \(s(N)\le j\), with the inputs specified in Supplement~S5 \cite{knuth,flammenkamp,cliftfour}. Let \(m\) be odd and \(k\ge1\). If \(s(2^{k-1}m)>4\) and \(s(2^km)=4\), then \(k\le\lambda(m)-8\). Equality holds for the following two disjoint sets:
\[
 \begin{array}{lll}
 m=83\cdot2^{k+2}+71, & k\ge9, & w(m)=8,\\
 m=1157\cdot2^{k-2}+67, & k=13\ \text{or}\ k\ge18, & w(m)=7 .
 \end{array}
\]
In either case the \emph{full} optimal length is
\[
 \ell(2^t m)=
 \begin{cases}
 k+t+13,&0\le t<k,\\
 k+t+12,&t\ge k .
 \end{cases}
\]
For every odd \(m\ne191\), \(K_4(m)\le\max\{0,\lambda(m)-8\}\), whereas \(K_4(191)=1\). Thus at every \(d\ge17\) the maximum of \(K_4(m)\) over odd \(2^d\le m<2^{d+1}\) is \(d-8\). This maximum concerns the profile \(\min\{5,s(2^tm)\}\); changes between values at least five are invisible.
\end{theorem}

Clift identified the \(83/71\) family, gave an upper construction and discussed unbounded delay and the candidate constant eight \cite{cliftdouble}. The contribution here is the all-parameter certificate of the transition, the sharp bound for an entry into level four, and the enumeration of its equality cases. Under the same completeness assumption, the first family has $k=\lambda(m)-8$ at arbitrarily large $k$. Thus at $q=1$, $a=4$ the coefficient of $\lambda(m)$ in Theorem~\ref{as:paths10} cannot be reduced below one with a uniformly bounded additive term.

\begin{proof}
Let \(p_0>\cdots>p_{w-1}=k\) be the bit positions of \(2^km\). Then \(p_0-2k=\lambda(m)-k\). Write \(C_j(p)\) for membership in the union of the catalogue rows through level \(j\). Under the hypothesis, a transition from above four to exactly four is equivalent to
\[
 C_4(p)\ \land\ \neg C_3(p)\ \land\ \neg C_4(p-\mathbf1).
\]
The first two tests place the target at level four. The last excludes its half even from that level. Thus some inequality of a row containing $p$ must fail when one is subtracted from every position. The complete boundary check (C1) in Supplement~S5 excludes all such transitions for \(p_0-2k\le7\). At equality \(p_0-2k=8\), the same check confines them to two lines in bit-position space. Substitution into the rows for \(p\) and \(p-\mathbf1\) gives the stated parameter ranges (S5); summing the corresponding powers of two gives the two odd bases.

On each line \(\lambda(m)=k+8\). The initial chains in S5 have length \(k+13\), hence excess at most five at every shift. Before the first entry this excess is at least five and at entry it equals four. A later drop below four would contradict constancy of \(\min(4,s(2^tm))\) for \(t\ge1\). Adding the top bit position \(k+8+t\) gives the asserted full lengths. Every other visible drop ending below four is confined to \(t=1\). For \(\lambda(m)\ge9\) this shift also satisfies the bound; among smaller odd bases the exact lower-level test gives just \(m=191\) with a positive delay (S5). Finally, for \(d\ge17\) put \(k=d-8\) in the first family. It realizes the upper bound.
\end{proof}

For \(k=9\), the first family gives \(m=170055\). Thus \(43\,534\,080\) and its double \(87\,068\,160\) have the same shortest addition-chain length, \(30\). At every \(d\ge26\) two odd bases attain the maximum; for \(17\le d\le25\) there is one, except at \(d=21\) where there are two.

\begin{table}[ht]
\centering
\caption{Largest possible stabilization time of the truncated profile, by binary weight of the odd length. An infinite supremum means unbounded finite delays.}\label{dmain:weight}
\begin{tabular}{c|cccccc}
\(w(m)\)&\(1\)--\(4\)&\(5\)&\(6\)--\(8\)&\(9\)--\(12\)&\(13\)--\(16\)&\(\ge17\)\\ \hline
\(\sup K_4(m)\)&0&1&\(\infty\)&3&1&0
\end{tabular}
\end{table}

Table~\ref{dmain:weight} describes a second distinction: delay is unbounded at each of three weights, but bounded again at higher weights because entry into the observed levels is more restricted. Explicit six- and seven-bit witnesses, with actual excess \(5\) before \(t=k\) and \(4\) thereafter, are
\[
 A_k=2^{2k+9}+2^{k+5}+2^{k+2}+21,\qquad
 B_k=2^{2k+8}+7\cdot2^{k+2}+21\qquad(k\ge6).
\]
For six ones any such transition requires \(\lambda(m)\ge2k+9\), and \(A_k\) attains equality. The eight-bit case is the first ray of Theorem~\ref{dmain:sharp}. At weights above sixteen the profile is identically five; this says nothing about drops between larger values of the full \(s\). The Supplement gives the remaining finite maxima, proofs for the extremal weights, and a complete determination of the full profiles of the 2,048 odd bases below \(4096\).

\section{Transfer of stabilization to string assembly indices}
\label{bridge:sec}
Suppose one symbol occurs $n$ times and another $2^h n$ times. Any addition sequence supplying both counts must first reach $n$ and then grow by a factor of $2^h$. The $h$ doublings that achieve this growth also supply all the intermediate powers of two times $n$. This gives a class of strings whose minimum assembly cost records the arithmetic cost of each doubling exactly.

We use the string-assembly model of \cite{rsos,jointspace}. Let $\Sigma$ be a finite alphabet of basic symbols and $\Sigma^+$ the set of nonempty finite strings over it. For $v\in\Sigma^+$, the number $q=|\operatorname{alph}(v)|$ counts the distinct symbols actually occurring in $v$, rather than the number of symbols available in the ambient alphabet. In the minimization below we may take $\Sigma=\{\sigma_1,\ldots,\sigma_q\}$ as the active alphabet.

An assembly space $\Omega=(V,E,\varphi)$ is a finite directed acyclic multigraph with sources $\Sigma$ and assembly step vertices $S_\Omega$, so $V=\Sigma\sqcup S_\Omega$. Each step has two incoming operand edges. The map $\varphi:E\to\{+1,-1\}$ records the left and right positions in the concatenation of their strings. Coincident operands give two parallel edges, counted separately. Vertices record formation histories and may carry equal strings. The size $|\Omega|=|S_\Omega|=|E|/2$ counts assembly steps. An assembly plan $\Omega_v$ retains the target $v$ and its ancestors, with unused symbols omitted; an optimal plan $\omega_v$ gives an assembly index $\operatorname{ASI}(v)=|\omega_v|$. Basic symbols cost zero. Previously formed strings may be reused, including twice in the same step, and copies incur no additional cost. For a single target, equal strings may be shared without increasing the number of steps. This is the reusable-string optimization used below; pathway depth and assembly-space energy are separate quantities.

Let $\operatorname{Par}(v)$ count the occurrences of $\sigma_1,\ldots,\sigma_q$ in $v$. Write $M(\mathbf n)$ for the minimum assembly index over all orders with the prescribed positive symbol counts $\mathbf n=(n_1,\ldots,n_q)$. Thus the minimum is over strings satisfying $\operatorname{Par}(v)=\mathbf n$. Transposition of an addition circuit \cite{olivos} gives
\begin{equation}\label{chain:wordcost}M(\mathbf n)=q-1+L(\{n_1,\ldots,n_q\}).\end{equation}
For a direct proof, prune an assembly plan with $r$ steps to its output. Give the output path count one and propagate counts back along both input edges of each step, counting parallel edges twice. At a vertex with $d$ outgoing edges, summing the $d$ received counts costs at most $d-1$ additions. There are $2r$ edges and $q+r-1$ vertices other than the output, all used. Thus at most $r-q+1$ additions produce the symbol counts. Conversely, take a shortest addition sequence with $u$ additions. Attach each symbol at the vertex for its required count and traverse the sequence backwards. At each vertex concatenate its attached symbols and incoming strings in any order, then send the result to both operand occurrences. All $u+1$ vertices receive something by minimality. The lists contain $q+2u$ pieces altogether and require $q+u-1$ assembly steps. Each symbol reaches the initial vertex once per dependency path from $1$ to its specified count, hence with its required multiplicity.
\begin{corollary}
A ladder of prescribed multiplicities.
\label{bridge:ladder}
Suppose the set $T$ of distinct multiplicities of $q$ symbols satisfies
$\{n,2^h n\}\subseteq T\subseteq\{n,2n,\ldots,2^h n\}$, with $n\ge1$ and $h\ge0$. Interior rungs are optional and any rung may occur for several symbols. Then
\[
                         M(\mathbf n)=q-1+h+\ell(n).
\]
\end{corollary}
\begin{proof}
Let $j$ and $k$ be the positions of $n$ and $2^h n$ in an increasing addition sequence. Then $j\ge\ell(n)$ and $2^h n\le 2^{k-j}n$ because the maximum can at most double at each step. Thus $k\ge\ell(n)+h$. A shortest chain for $n$ followed by $h$ doublings reaches every required rung at that cost, including when $h=0$. Apply~\eqref{chain:wordcost}.
\end{proof}
To build a string attaining the bound, divide all counts by $n$. On the ladder of prescribed multiplicities $1,2,\ldots,2^h$, attach an output for each symbol to the rung bearing its reduced count. Transposition gives a motif in $q-1+h$ assembly steps. Raising the motif to its $n$th power costs another $\ell(n)$ assembly steps. Interior rungs are available as soon as both endpoints are required. Each additional symbol still contributes one assembly step through $q-1$.

For a fixed vector \(\mathbf n\) satisfying Corollary~\ref{bridge:ladder}, let \(M_t=M(2^t\mathbf n)\), doubling every multiplicity at each shift. Then
\[
 M_t-t=(q-1+h)+b_n(t).
\]
Thus the first shift from which every doubling costs one additional assembly step is \(K(n)\), and Corollary~\ref{dmain:stabilitytest} tests stability of the string assembly costs from the outset. This transfer holds independently of the catalogues.

\begin{example}
Two optimal strings whose lengths differ by a factor of two.
For $q=3$ and counts $(n,2n,4n)$ a suitable motif is $P=a(cbc)^2$. The four assembly steps are $U=cb$, $V=Uc$, $Y=VV$, $P=aY$. Thus $P^n$ attains $M(n,2n,4n)=4+\ell(n)$. Put $m=170055=83\cdot2^{11}+71$. Under the catalogue hypothesis of Theorem~\ref{dmain:sharp}, the level-four classification gives $\ell(2^8m)=\ell(2^9m)=30$ (the construction of the ray is due to Clift \cite{cliftdouble}). Consequently the two strings $P^{2^8m}$ and $P^{2^9m}$ have lengths $304\,738\,560$ and $609\,477\,120$, respectively, and each has assembly index $34$.
\end{example}

Under the catalogue hypothesis of Theorem~\ref{dmain:sharp}, the same calculation applies to every $m_k=83\cdot2^{k+2}+71$, $k\ge9$. In a fixed ladder with $n=2^t m_k$, going from $t=k-1$ to $t=k$ doubles every count at the same minimum cost. Every other doubling costs one extra assembly step. Changing $q$ or $h$ changes the baseline $q-1+h$ and leaves the position of this saving at $k$.

\paragraph{}
Arbitrary prescribed counts and block order.\\
The ladder formula is a closed-form special case of a more general transfer. Let $\mathbf n=(n_1,\ldots,n_q)$ have positive entries and let $T=\{n_1,\ldots,n_q\}$. Here $q$ counts symbols and $h=|T|$ counts distinct numerical multiplicities. Repeated multiplicities occur once in $T$, but once per symbol in sums over $i$.

\begin{corollary}
\label{v13:fixed-composition}
For every prescribed positive count vector,
\[
 M(2^t\mathbf n)=q-1+t+L_{\rm st}(T)\qquad(t\ge K_T).
\]
For pairwise distinct symbols $\sigma_1,\ldots,\sigma_q$, define
\[
 \Delta_{\mathbf n}(t)=
 \operatorname{ASI}(\sigma_1^{2^tn_1}\cdots\sigma_q^{2^tn_q})-M(2^t\mathbf n).
\]
For all $t\ge\max\{K_T,K(n_1),\ldots,K(n_q)\}$ one has
\[
 \Delta_{\mathbf n}(t)
 =(q-1)t+\sum_{i=1}^{q}\ell_{\rm st}(n_i)-L_{\rm st}(T).
\]
The scalar and joint horizons give a computable upper bound for the onset of this linear formula.
\end{corollary}

The first identity follows from \eqref{chain:wordcost}. Supplement~S7.2 proves the block identity
$\operatorname{ASI}(\sigma_1^{n_1}\cdots\sigma_q^{n_q})=q-1+\sum_i\ell(n_i)$
and the corollary. Thus for $q\ge2$ the penalty grows eventually by $q-1$ at each doubling, at every fixed positive frequency vector. The empirical single-symbol entropy is constant along these scales; the statement concerns neither block entropy nor an entropy rate.

\begin{example}
A prescribed composition outside a dyadic ladder.
\label{v13:joint-example}
For $T=\{7,191\}$, eleven-step sequences supply $T$ and $2T$. The scalar costs and the shift-invariant lower bound for the target $191$ with seven binary ones give $L_{\rm st}(T)=10$ and $K_T=1$ (Supplement~S7.2). Consequently
\[
 M(7,191)=12,\qquad M(2^t7,2^t191)=11+t,\qquad
 \Delta_{(7,191)}(t)=t+4\quad(t\ge1).
\]
By contrast $F_2(2^t198)=9+t$ for every $t\ge0$, so $K_2^{\rm str}(198)=0$. Fixed-composition and free-composition stabilization indices need not coincide.
\end{example}

\section{Discussion and scope}
A finite computation can settle a question about infinitely many doublings in two ways. One can calculate a horizon from the input and determine the cost there. Alternatively, a lower bound valid at every shift can meet the cost of a single exhibited plan. For $191$ the latter argument fits in a few lines in Section~\ref{chain:paths}. For other inputs the horizon guarantees termination even when such a certificate is unavailable. Supplement~S6 gives the tests behind the $4\,984$ scalar certificates for the $8\,192$ odd bases below $2^{14}$. The sharper $d-8$ conclusion concerns entries into level four and uses the completeness of the small-step catalogues specified in Theorem~\ref{dmain:sharp}. A reader can check the general horizon without assuming their completeness.

The prescribed-target extension is independent of the catalogue premise. Its improvement uses distinct endpoints and does not alter the scalar estimate when only one target is prescribed. Supplement~S7.3 shows that, in the range stated there, the leafwise carry loss is already sharp when only the value, cap and leaf count are known; improving it within that range needs additional structural information.

Supplement~S6.6 gives portable certificates of both the minimum normalized cost and the earliest shift for one fixed run pattern. Checking a feasible integer schedule, a dual flow and a tight path does not repeat the optimization. The classical assignment routine uses $O(f^3)$ arithmetic operations for $f$ supply--demand pairs, while the outer enumeration of topologies and offsets remains. A local certificate does not establish that this enumeration is complete.

Two further questions delimit the scalar results. Does the sharp constant for first entry into level four extend to higher destination levels as proposed by Clift \cite{cliftdouble}? For unrestricted single targets, can one certify a lower bound in polynomial time, or give a hardness reduction that forces the required intermediate values starting from $1$? The bounded-excess algorithm exploits the small number of departures from doubling. The hardness result in Supplement~S3 allows the input to prescribe a prefix. The complexity of the unrestricted single-target problem remains open.

\paragraph{Data availability statement.}
The supplement gives the full specifications of the computations. The accompanying \path{ASI_Verification_Lite.zip} supplies selected verification programs, scalar stability certificates and coverage tests for the two equality families. Its \path{extensions/v13} directory adds the fixed-target checks, explicit word constructions and portable network certificates described in S6.6 and S7. Verification of the conditional catalogue result uses the versions of Clift's external catalogues identified in Supplement~S5 \cite{cliftsmall,cliftfour}.

\paragraph{Use of AI.}
OpenAI's ChatGPT was used to develop verification code and examine computations.

\paragraph{Funding:} This research received no external funding.

\paragraph{Acknowledgements:} The author thanks Szymon Łukaszyk, Wawrzyniec Bieniawski, and Szymon Tworz for their clarifications, formal corrections, improvements, and meticulous revisions, and Mariola Bala for maintaining order and discipline within our Fellowship of the Ring.

\paragraph{Conflicts of Interest:} The author declares no conflict of interest.

\end{document}


\title{Supplementary Information\texorpdfstring{\\[.25em]}{ — }\paperTitle}
\author{Piotr Masierak\,\protect\orcidlink{0009-0008-9895-5329}}
\thanks{\href{mailto:pmasierak@patent.pl}{pmasierak@patent.pl}}
\affiliation{Dom Zatorski Scientific Foundation, Chorzów, Poland}
\date[]{November 26, 2026}
\begin{abstract}
Sections S1--S3 give the sparse-input algorithm, certificates for exact lengths and the complexity of continuing a prescribed prefix. Section S1.1 includes the integral formulation for compressed chains, its proof and the bit-complexity analysis. Section S4 proves an effective first-entry bound by shortening a run of doublings. Section S5 states the catalogue-completeness assumption used for the sharp level-four profile and distinguishes that assumption from the finite arithmetic checks. Section S6 strengthens the effective bound by counting doublings on disjoint paths and develops certificates of stabilization, including portable certificates for a fixed run pattern. Section S7 gives an effective horizon for any prescribed finite target set, proves the corresponding fixed-composition and block-order results, and identifies a sharpness limit of the leafwise carry estimate.
\end{abstract}
\maketitle
\setcounter{tocdepth}{1}
{\small\tableofcontents}
\clearpage

\section{Algorithms for one target and for all targets up to a limit}\label{sec:si-algorithms}
An addition chain for $n$ starts at $1$, ends at $n$ and increases strictly. Each later term is the sum of two earlier terms, which may coincide. Its length counts additions, with $\ell(1)=0$. We write $\ell(n)$ for the minimum length, $\lambda(n)=\lfloor\log_2 n\rfloor$, $w(n)$ for the binary weight and $s(n)=\ell(n)-\lambda(n)$ for the optimal excess. A small step leaves $\lambda$ unchanged relative to the preceding term. An immediate doubling replaces the current maximum by twice that value. A step other than an immediate doubling may itself double an older term. Thus small steps and steps other than immediate doublings are different classifications; their counts need not agree.

\subsection{Contracting immediate doublings}
For an input budget $k$, put $D=1+\lambda(n)$, $w=w(n)$ and $\sigma=k-\lambda(n)$. Reject $k<\lambda(n)$. For $k\ge\lambda(n)+w-1$ the binary method supplies an immediate witness, recording the gaps between occupied positions as doubling runs for sparse input. In the remaining cases let $f$ count the steps other than immediate doublings in a chain of length $j$. Then
\[
 a_j\le 2^{j-f}\varphi^f,\qquad \varphi=(1+\sqrt5)/2.
\]
Inductively bound $(a_i,a_{i-1})$ coordinatewise by a multiple of $(1,1/\varphi)$, starting at $(1,0)$. Doubling increases the multiplier by at most two. Otherwise $a_i\le a_{i-1}+a_{i-2}$, so the multiplier increases by at most $\varphi$. For $a_j=n$ and $j\le k$ this gives $f\log_2(2/\varphi)\le j-\log_2 n\le \sigma$. Since $\varphi^4<8$, one has $\log_2(2/\varphi)>1/4$ and $f\le4\sigma$, also at $\sigma=0$ \cite{knuth,clift}.

Contract an increasing optimal chain into runs of $h_i\ge0$ doublings with bases $v_i$. For $1\le j\le f$ we have
\[
 v_0=1,\quad v_j=2^{u_j}v_{p_j}+2^{z_j}v_{q_j},\quad
 p_j,q_j<j,\quad 0\le u_j\le h_{p_j},\quad 0\le z_j\le h_{q_j},\qquad
 n=2^{h_f}v_f .
\]
Choose the first incoming edge as primary. Put $\rho_0=0$, $\rho_j=\rho_{p_j}+u_j$, $E_i=\rho_i+h_i$, $d_j=\rho_{q_j}+z_j-\rho_j$, and $c_j=2^{-\rho_j}v_j$. Then
\[
 c_0=1,\quad c_j=c_{p_j}+2^{d_j}c_{q_j},\qquad n=2^{E_f}c_f .
\]
In a shortest chain every gate is an ancestor of the output for any fixed justification of its sums: otherwise delete all values outside that ancestor set. In particular the topology can be pruned in advance.

The expanded output has at most $2^f$ leaves of the form $2^b$ with $b\ge0$. In reducing the multiset of powers to binary form each carry removes one item. Every leaf exponent lies at most $2^f-1$ positions below its final set bit. Writing $\operatorname{supp}_2(n)$ for the occupied positions, all leaf exponents therefore belong to $C_f(n)=\{b-i:b\in\operatorname{supp}_2(n),\ 0\le i<2^f,\ i\le b\}$. If $w>2^f$ discard the topology; otherwise $|C_f(n)|\le 2^{2f}$. The all-primary leaf has exponent $E_f$. At gate $j$ choose any path from $j$ to the output. Extend backwards along the primary paths in its two parents, one through each input edge of $j$. The exponents of the resulting leaves differ by $d_j$, since all subsequent shifts are shared. Thus $d_j\in C_f(n)-C_f(n)$. This is a relative shift and may be negative.

\begin{lemma}[Integral schedules]\label{sa:exact-schedule}
For fixed parents and shifts, compute $c_j$ and $e=\lambda(n)-\lfloor\log_2c_f\rfloor$. If $2^ec_f\ne n$, the assignment is impossible. Otherwise an increasing chain with this description and cost at most $k$ exists if and only if the following linear program has a feasible point of objective at most $k$:
\[\begin{aligned}
 &\min \quad f+\sum_{i=0}^f(E_i-\rho_i),\\
 &\rho_0=0,\quad 0\le \rho_i\le E_i\le k,\quad E_f=e,\\
 &\rho_{p_j}\le \rho_j\le E_{p_j},\qquad
   \rho_{q_j}\le \rho_j+d_j\le E_{q_j},\\
 &E_{j-1}+1+\lfloor\log_2(c_{j-1}/c_j)\rfloor\le\rho_j
               \qquad (1\le j\le f).
\end{aligned}\]
\end{lemma}
\begin{proof}
In a chain, $E_i$ counts doublings on a single path, so $E_i\le k$. The two interval constraints place the operands in available runs. The last inequality is the requirement $2^{E_{j-1}}c_{j-1}<2^{\rho_j}c_j$ that the gate increase the previous maximum. This gives a feasible point of the stated cost.

Conversely the constraints are differences with integral right-hand sides. Each column of the transposed coefficient matrix has at most two nonzero entries, of opposite signs if both occur. Expanding a square minor along a column with at most one entry, or observing that the sum of its rows is the zero vector when each column has two opposite entries, proves total unimodularity. The bounded nonempty polyhedron therefore has an integral optimal vertex. From that vertex, successively take the indicated two operands, add them and perform the next $E_j-\rho_j$ doublings, starting with the initial run from $1$. The result is an increasing chain ending at $n$, of the objective cost, even when a gate itself is a doubling.
\end{proof}
An optimal vertex can be obtained in polynomial time by rational linear programming. The implementation uses difference constraints and shortest augmenting paths. For $f=0$ one simply tests a power of two.

There are at most $\prod_{j=1}^f j^2$ ordered parent assignments and at most $2^{4f^2}$ difference assignments. In dense arithmetic each normalized dyadic has numerator and denominator of $O(fD+f+1)$ bits (multiply by a common power of two if necessary). For sparse input, store its canonical list of exponent positions, allowing signed positions in normalized dyadics. Each $c_j$ is a sum of at most $2^j$ powers with absolute initial exponent at most $j(D-1)$. Merging equal exponents and carrying uses at most $2^j-1$ merges; all positions fit in $O(\log(D+1)+f+\log(f+1))$ bits. To compute $\lfloor\log_2(x/y)\rfloor$ for two positive lists, let $e$ be the difference of their leading exponents and compare $x$ to $2^ey$ lexicographically, from the largest differing position. Return $e$ or $e-1$. Thus the per-assignment cost in sparse arithmetic is $2^{O(f)}\operatorname{poly}(w\log(D+1)+\log(k+2))$. Constructing $C_f$ and the graph has the same allowed dependence. Returning the schedule and its gate parents is sufficient to specify a witness, even if writing down every doubled integer would be prohibitive. The dense solver used in the research calculations was checked on all $63$ targets $1,\ldots,63$ and eight budgets $0,\ldots,7$ in four modes (unrestricted; at most $0$, $1$, or $2$ exceptions). All $2{,}016$ decisions matched a separate exhaustive search. The sparse companion was separately checked on the $504$ unrestricted oracle decisions, the stabilization of $15$, and both adjacent budgets $10^9+4,10^9+5$ for $15\cdot 2^{10^9}$. It returns a run-length witness for the higher budget and rejects the lower one without constructing a billion-bit integer.

\subsection{Computing stability from sparse data}
Write $m=\lambda(n)$, $F=4\sigma$, $U=2^F-1$, $H=(2F+1)(m+U)+U$. The finite-horizon theorem in the main text gives an exact test of the existence of a shift satisfying the budget $m+t+\sigma$, by testing just $t=H$. Expressing $H$ in binary costs $O(\log(m+2)+\sigma+\log(\sigma+1))$ bits. The support of $2^t n$ is obtained by adding $t$ to each input position; the budget excess is still $\sigma$. Consequently a threshold test, or binary search for its first success on $[0,H]$, takes $2^{O((\sigma+1)^2)}\operatorname{poly}(w\log(D+1))$ bit operations (the loop uses $O(\log(H+2))$ calls). Search $\sigma=0,\ldots,w-1$. Its first successful value $s_*$ gives $\ell_{\rm st}(n)=m+s_*$; its first successful shift is precisely the stabilization index $K(n)$. The resulting dependence on $w$ is $2^{O(w^2)}$ times a polynomial in the sparse input length. This algorithm uses the horizon as a stopping rule, not as a prediction of the typical stabilization index.

\subsection{An explicit table with almost linear cost in its output length}
For each $1\le n\le X$ initialize a proven upper bound by the $2^b$-ary method: build all digits up to $2^b-1$ by increments, and read the digits of $n$ with at most $b$ doublings and one addition per remaining block. If the precomputed digits and subsequent steps interleave in value, discard unused values and sort, which cannot increase cost. Put $m=\lfloor\log_2X\rfloor$. With $b=\max(1,\lfloor\log_2(m+2)-2\log_2\log_2(m+2)\rfloor)$ (taking $b=1$ at the small endpoint), uniform initial excess over $\lambda(n)$ is at most
\[
 R\le (1+o(1))\,m/\log_2m.
\]
Indeed $2^b=o(m/\log m)$ and $b\sim\log_2 m$. The leading digit is already available. The remaining blocks cost at most $\lambda(n)$ doublings in total and at most $\lfloor\lambda(n)/b\rfloor$ digit additions. Thus $R=2^b+\lfloor m/b\rfloor$ is a valid uniform choice. One can also choose $b$ separately at each bit length.

Enumerate all increasing value sequences with last value $\le X$ and excess at most $R$; update the minimum for each endpoint. Exhaust all distinct admissible next sums, without merging different histories ending at the same number. Excess is nondecreasing along any chain, so every optimum is covered. A prefix of $j$ operations ending at $x$ extends by $m-\lambda(x)$ doublings into $[2^m,2^{m+1})$, keeping its excess $a=j-\lambda(x)$. For fixed $j$ this map is injective (truncate after $j$). Thus the prefix count is at most $(m+R+1)\sum_{a\le R}G(m,a)$, where $G$ counts \emph{chains} of $m+a$ operations ending in this dyadic interval. For $a\le m^{0.9}$ the elementary bound $f\le4a$ gives $\exp(O((a+1)\log m))$ via choices of exceptional positions and parents. Above that threshold use the preprint bound $G(m,a)\le (C m^2/a)^a$ \cite[Theorem~1]{dkdv}. The latter expression increases over this range for large $m$. At $a\le (1+o(1))m/\log_2m$ its logarithm is at most $(\log 2+o(1))m$ (logs here natural), so the total is $X^{1+o(1)}$. Polynomial factors in $m$ for enumerating the next sums and comparing integers are absorbed. This proof relies on the cited chain-count estimate, not on the corresponding bound for the number of endpoints. The reference run for $X=255$ improved constructive initial bounds by enumerating $505{,}461$ prefixes at excess at most $3$; a separate search agreed on every one of the $255$ lengths, including $\ell(255)=10$.

\section{Checkable lower bounds and numerical verification}\label{sec:si-cert}
For a finite set $T$ of positive targets, $L(T)$ is the fewest additions in one addition sequence containing every element of $T$. Such a computation is also called a joint addition chain; its increasing list is an addition chain for its largest entry. We have $L(T)=0$ for $T\subseteq\{1\}$. Otherwise, with $m=\max T$,
\[
 L(T)=1+\min_{1\le u\le\lfloor m/2\rfloor}L\bigl((T\setminus\{m\})\cup\{u,m-u\}\bigr).
\]
A shortest addition sequence may end at $m$. Removing its final addition gives a feasible addition sequence for one set in the minimum. Conversely, a sequence ending at the largest target in such a set can be extended by $m=u+(m-u)$.

A chain witnesses an upper bound. To certify equality, one also needs to exclude \emph{every} chain at the preceding budget. The backward recurrence for $L(T)$ yields one way to do this: at a node $(T,r)$ the verifier checks that the split of $\max T$ into two positive terms is covered for \emph{all} choices. Identical full states can share subproofs. For sorted required values $1=b_0<b_1<\cdots<b_j$, the lower bound
\[
        L(T)\ \ge\ \sum_{i=1}^{j}\left\lceil\log_2(b_i/b_{i-1})\right\rceil
\]
follows by counting doublings possible between consecutive required values. Forward search can instead close a branch by testing its last few additions exactly. These are complementary verification methods; neither supplies a polynomial bound on certificate size for arbitrary binary targets.

\paragraph{Backward certificates.}
Normalize a target set by dropping $1$. A certificate for $L(T)>r$ specifies arithmetic closures or a disjoint, exhaustive cover of $u=1,\ldots,\lfloor\max(T)/2\rfloor$. At a singleton split the checker reconstructs the child and requires a reference to an earlier node with that child state and budget $r-1$. For an interval, order the affine functions $1$, the retained targets, $u$ and $\max(T)-u$ identically at both endpoints. The order then holds throughout. A ratio of adjacent positive affine functions is monotone or constant, so the smaller endpoint ceiling of its binary logarithm bounds its contribution to the successive-target growth bound everywhere. If the sum of those minima exceeds $r-1$, the entire interval closes. Ties contribute zero. If the budget equals the number of required nonunit values, those values must themselves form a legal chain with $1$. These checks are polynomial in the size of the explicit proof and binary input; interval lengths need not be enumerated. The backward checker was tested on $900$ decisions ($1\le N\le100$, $0\le k\le8$): $664$ NO files and $236$ YES witnesses agreed with direct search, and all ten certificate mutations in the test suite were rejected.

\paragraph{Forward arithmetic certificates.}
Unfolding $r$ additions expresses $N$ as a sum of at most $T=2^r$ old terms, with repetition. If $r\ge2$ and $2^{r-1}\nmid N$, one may take $T=3\cdot2^{r-2}$. Give old terms weight $1$ and each gate the sum of its input weights. With fewer than $r$ gates the weight is at most $2^{r-1}$. If each step after the first doubles the preceding gate, the output is divisible by $2^{r-1}$. Otherwise at the first departure, step $j\ge2$, only the preceding gate has weight $2^{j-1}$ and all other values have weight at most $2^{j-2}$. The new weight is at most $3\cdot2^{j-2}$. Each of the remaining $r-j$ steps can at most double the maximum, giving the bound.

For $B\ge\max A$ and a modulus $q$, provide $h$ with $h(0)=0$ and
\[
 h((t+a)\bmod q)\le h(t)+B-a \quad(a\in A,\ t\bmod q).
\]
Summing along a representation gives $N+h(N\bmod q)\le BT$. Its violation proves impossibility. One can store only values below a common cap, since $B-a\ge0$ and inequalities out of capped residues hold automatically. For $A=(1,2,3,6,12,24)$, $r=2$, $N=83$, use $B=q=24$, $h(0)=0$, $h(12)=12$, and $14$ elsewhere: $83+14>96$. Growth alone gives no contradiction. Conversely, for $A=(1,2,3,4)$ the identity $23=5\cdot4+3$ meets the improved six-leaf bound for $r=3$, but its three-gate suffix does not exist. The first new value $x\le8$, the second $y\le16$; $2y$ is even, and $y+\mathrm{old}\le20$. For $y+x>20$ one needs $y=2x$, giving a multiple of three. Unused or independently computed intermediates cannot improve these bounds. Four gates suffice via $7,8,15,23$. Scalar leaf counts lose information about shared gates.

\paragraph{Local dual certificates.}
For a fixed description in S1.1, write the constraints as $y_v-y_u\le c_{uv}$. Summing around a negative cycle proves infeasibility. For a nonempty schedule polytope, linear-programming duality gives
\[
 \min_y\sum_i(E_i-\rho_i)=-\min_F\sum_{uv}c_{uv}F_{uv}.
\]
Here $F\ge0$ has net outflow $+1$ at each $E_i$, $-1$ at each $\rho_i$ and zero elsewhere. Multiplying the inequalities by $F_{uv}$ and adding gives the displayed sign. The incidence matrix and the integral balances admit an integral optimal flow. Its cycles can be removed because a feasible difference system has no negative cycle. At most $f+1$ unit paths remain. Thus each rejected case supplies a cycle, a flow proving the cost exceeds the budget, or an arithmetic rejection. Enumerating the cases as in S1.1 bounds the size and checking time of the entire explicit certificate by $2^{O((\sigma+1)^2)}\operatorname{poly}(D)$.

\paragraph{Multipliers and all subsequent doublings.}
The classifications at small excess \cite{knuth,altman} imply $\ell(n)\ge B_0(n)=\lambda(n)+b(w(n))$, where $b(1)=0$, $b(2)=1$, $b(3)=b(4)=2$, and $b(w)=3$ for $w\ge5$. Concatenating a chain for $n$ with a chain for $q$ scaled by $n$ gives $\ell(n)\ge\ell(qn)-\ell(q)$. In particular $B_*(n)=\max(B_0(n),B_0(3n)-2)$ satisfies
\[
     \ell(2^j n)\ge B_*(2^j n)=j+B_*(n).
\]
If equality is witnessed at a single shift, appending doublings proves equality for every later shift. For $b\ge2$, $a\ge b+3$ and $t\ge0$ put $n=2^t(2^a+2^{a-1}+2^b+1)$. The bits of $3n$ have distinct positions $t+a+2,t+a-1,t+b+1,t+b,t+1,t$, so $\ell(n)\ge t+a+3$; the binary method attains it. For $(a,b,t)=(100000,33333,0)$ this gives $\ell(n)=100003$. These families are consequences of the classical low-excess lower bounds, requiring no new exhaustive classification. The test cannot prove a lower bound above $\lambda(n)+3$ even with arbitrary multipliers: $\lambda(qn)\le\lambda(n)+\lceil\log_2q\rceil$ and $\ell(q)\ge\lceil\log_2q\rceil$. In $1,\ldots,255$, the bound $B_0$ is exact in $217$ cases and $B_*$ in $237$ ($113$ distinct odd parts in the latter). For example, $B_*(191)=10$, and equality is attained at shift one: $\ell(382)=11$. Independent rejection at budget $10$ for $191$ establishes $\ell(191)=11$ and $K(191)=1$.

\paragraph{Three final additions in closed form.}
For a legal prefix ending at $M$, let $P=A+A$, $B=\{p\in P:p>M\}$. Here $j\cdot B$ means dilation, not a sumset. The cumulative reachable sets for $N>M$ are obtained by adding the following terms at each stage:
\[
\begin{array}{c|l}
  1&B\\
  2&B+A,\quad 2\cdot B\\
  3&B+P,\quad (2\cdot B)+A,\quad 2\cdot(B+A),\quad 3\cdot B,\quad 4\cdot B .
\end{array}
\]
If the first useful result is $x\in B$ and the second depends on it, then the second is $x+c$ or $2x$, and the final step adds an old value, adds $x$, or doubles the second. If two independent new results are used their sum lies in $B+P$. Conversely, a value $x+(c+d)\in B+P$ can always be scheduled as $(x+c)+d$, so no illegal computation of $c+d\le M$ is implied. The other terms have immediate witnesses. Sorted pair sums and complement lookup give $O(h^2\log(h+1))$ comparisons/arithmetic operations on $O(\log N)$-bit numbers. For example, test $B+P$ by looping over $x\in B$ and looking up $N-x$ in the sorted $P$. The other sums use the same lookup after the indicated dilations.

\section{Prescribed prefixes: exact algorithm and reductions}\label{sec:si-prefix}
The decision problem \textsf{Prefix-Ext} takes an explicitly supplied legal increasing chain $A=(1=a_0<\cdots<a_{h-1}=M)$ as a mandatory prefix. It asks whether at most $r$ additions can be appended to reach $N>M$. The earlier operations have already been performed, and their results and order are fixed.

\begin{theorem}[Suffix test and matching conditional exponent]\label{thm:alg-prefix}
For $D=1+\lfloor\log_2N\rfloor$, \textsf{Prefix-Ext} admits a deterministic witness-producing algorithm in
\[
  2^{O(r\log(r+1))}(h+1)^{\lceil(r+1)/2\rceil}
          \operatorname{poly}(r,D,\log(h+1))
\]
bit operations. With $r$ part of the input it is NP-complete and is W[1]-hard parameterized by $r$, even when $M<N<2M$. For each fixed $r\ge2$, under the coloured $(r+1)$-SUM hypothesis \cite{gokaj}, there is no word-RAM algorithm with time $O(h^{\lceil(r+1)/2\rceil-\epsilon}\operatorname{poly}(\log N))$ for any $\epsilon>0$, even for $O_r(\log h)$-bit entries.
\end{theorem}

\begin{proof}
For $j=1,\ldots,r$ enumerate every topologically ordered binary addition graph on $j$ new gates. At each gate an argument is a previous gate or a fresh port representing an unconstrained choice in $A$; the same old value may fill arbitrarily many ports. Remove graphs with a gate that is not an ancestor of the output. A port's positive integer coefficient in the output is its number of paths to the output, counting parallel edges separately. The two ports of any all-old gate have the same coefficient and must satisfy $a+b>M$. These are the only inequalities needed: all other gates have an all-old ancestor and add positive values. A solution of this weighted equation gives a legal suffix by sorting the numerical gate values and deleting duplicates. Both parents of a retained value are strictly smaller. An actual suffix gives one of the enumerated assignments by pruning unused gates.

At most $j+1$ ports are old, since at least $j-1$ of the $2j$ inputs link new gates. Order the disjoint constrained pairs consecutively, and divide all ports into two halves whose sizes differ by at most one. Each side filters assignments violating an internal pair constraint. For each attainable weighted sum retain one assignment; if a pair was split retain the assignment with the greatest crossing-port value. Matching complementary sums by sorting is sufficient, with the extra check that the two recorded maxima sum to more than $M$ for a split pair: the two assignments can be chosen independently. At most $\sum_{j\le r}(j!)^2$ ordered topologies are needed using ordered parents and an old-port symbol. All coefficients have $O(r)$ bits. This proves the time bound in Theorem~\ref{thm:alg-prefix}; at $r=0$ the decision is immediate. For example, omitting the $>M$ condition would admit $12=10+2$, then $24=12+12$ from $(1,2,3,5,10,15)$. The first addition is illegal, and the only legal first values below $24$ are $16,17,18,20$, none of whose complements is old.

\paragraph{Reduction from selection in lists.}
Suppose $k\ge2$ nonempty lists $Y_i$ of nonnegative integers are given and one number from each must sum to $T$. Put $S=\sum_i\max Y_i$, assume $0\le T\le S$ (otherwise reject), and set
\[
 U=S+1,\quad L=U+\max_i\max Y_i,\quad
 2^{P-1}>\max\{(k+1)L,kU+S\},\quad H_i=2^{P+i-1}.
\]
Use the smallest admissible $P$, or increase it by any fixed constant. The prefix is the increasing union of: every power of two through $H_k$; the binary chains for all $U+y$ ($y\in\bigcup_iY_i$); and all candidates $H_i+U+y$ for $y\in Y_i$. Powers are obtained by doubling, the small values have their binary chains, and each candidate is a sum of the two included terms $H_i$ and $U+y$. This is a legal prefix. It has polynomial encoded size. Set
\[
     B=\sum_i H_i,\qquad W=kU+T,\qquad N=B+W,\qquad r=k-1 .
\]
Since $H_1>L$, $H_k<M\le H_k+L$, $W<H_1$, and $B=2H_k-H_1$, we have $M<N<2M$. For a valid choice, first add the candidate in the highest class to any other selected candidate. The result exceeds $H_k+L$, and all subsequent sums increase, giving a legal suffix.

Conversely, as $N\le2M+1$, a sum of two new values overshoots the target. In its ancestor graph each gate after the first therefore adds one old term to the previous result. Thus $N$ is a sum of at most $k$ old terms. Split a candidate into high part $H_i$ and low part $U+y$. Treat pure powers as high and all other small terms as low. The sum $Q$ of the low parts and $W$ both lie in $[0,2^{P-1})$. The integer $B+W-Q$ is a sum of at most $k$ powers. If $Q<W$ its binary expansion keeps the $k$ bits of $B$ and gains at least one lower bit. If $Q>W$, the lowest bit of $B$ is replaced by $2^P-(Q-W)\in(2^{P-1},2^P)$, which has at least two bits. Both alternatives require more than $k$ powers. Hence $Q=W$. As $B$ has $k$ bits, every tag occurs once, without a carry. Even one bare tag would give $Q\le(k-1)U+S<kU$. Thus all terms are candidates, and subtracting $kU$ gives a selection summing to $T$.

\paragraph{Complexity consequences.}
Taking $Y_i=\{0,x_i\}$ proves NP-hardness by binary \textsf{Subset Sum} \cite{karp}. For membership in NP even if the budget is binary, any legal prefix with endpoint $M<N$ can reach $N$ in $O(\log N)$ new steps: double up to the largest $M'=2^jM\le N$, then add terms from this augmented prefix greedily to make up $N-M'$. Successive entries of the augmented prefix have ratio at most two, so the largest entry at most a positive remainder is at least half that remainder. Thus a yes instance has a polynomial witness: take a shortest continuation, whose length is bounded by both the given budget and this universal construction. One can check legality of the input prefix as well as the appended suffix in polynomial time.

For W[1]-hardness use \textsf{Clique} parameterized by the requested number $t$ of vertices \cite{downeyfellows}. Give each of $t$ positions a vertex list and each unordered pair of positions a list of edges, with both orientations. There are $k=t+\binom t2$ lists. Give vertex identifiers $1,\ldots,d$, put $C=d+1$, and use radix $2C+1$. The coordinates are ordered incidences $(i,j)$ for $i\ne j$. A vertex choice at position $i$ writes its identifier in every $(i,j)$; the edge choice for $i,j$ writes $C$ minus the identifier of its proposed endpoint in each corresponding coordinate. All other digits are zero. Target digit $C$ everywhere enforces endpoint agreement: each coordinate receives two contributions whose sum is below the radix. A solution is thus a clique in the simple graph. The numbers use $O(t^2\log(d+2))$ bits, with at most $td+2\binom t2|E|$ choices. The composed reduction gives $r=t+\binom t2-1$. This proves hardness under parameterized reductions; no claim about W[1] membership is required.

For fixed $k\ge3$ use the coloured $k$-SUM hypothesis in the word-RAM model \cite{gokaj}: choosing one integer from each of $k$ lists of at most $n$ polynomially bounded integers to sum to zero requires $n^{\lceil k/2\rceil-o(1)}$ time in the sense of excluding any fixed exponent improvement. Shift every entry by the same constant large enough to make it nonnegative and adjust the target by $k$ times that constant. The above prefix has $h=O_k(n\log n)$ entries of $O_k(\log n)$ bits; constructing it with sorting takes $n\operatorname{polylog}n$ time. With distinct entries per list and $n$ the largest list size, $h\ge n$ too. An $h$-exponent improvement $\epsilon>0$ at $r=k-1$ would give an $n$-exponent improvement $\epsilon'>0$ for this decision problem. All claims refer to arbitrary legal prefixes; the reduction does not work by simply setting $A=(1)$.\end{proof}

\paragraph{Near powers and a tractable special case.}
NP-hardness survives the additional requirements that $\max A$ be a power of two, that all smaller powers of two be present, and that the relative distance to the target be arbitrarily small (fixed beforehand). Extend the prefix by powers up to a new power $D_0>2H_k$, chosen so $N<\eta D_0$ for a fixed $\eta\in(0,1]$. Use target $D_0+N$ and budget $k$. A valid selection can be added to $D_0$ one term at a time. Conversely repeat the bit count with $k+1$ tags, the new highest tag being $D_0$; the choice of $P$ bounds all $k+1$ low parts. A carry-free choice of that highest tag must be the bare power $D_0$, leaving precisely the original list choices. For the variable-$k$ binary \textsf{Subset Sum} reduction one may also shift all tag positions by $J=z^c$, $z$ the source encoding length, keeping the full ladder. At most $O(z^2)$ included values are not powers (binary traces of the $U+y$ and the candidates). Taking an integer $c>\max(2,2/\epsilon)$ gives prefix excess over the growth bound $O(D^\epsilon)$ and $r=O(D^\epsilon)$ with $D$ the target bit length, for every fixed $\epsilon>0$. Writing down the padded prefix still costs polynomially many bits in $z$. This padding statement, with growing input integers, is separate from the fine-grained reduction.

In contrast, for the pure ladder $A=\{1,2,4,\ldots,M\}$ and $M<N<2M$, the minimum suffix cost equals $w(N)-1$: the path needs at least $w(N)$ old terms, and the nonleading bits can all be added to $M$. More generally allow $d$ nonpower extras $e_1,\ldots,e_d$ below $M$. Put $B'=(A+A)\cap(M,N]$. The cost is
\[
 1+\min_{b,\mathbf z}
 \left\{\sum_i z_i+w\!\left(N-b-\sum_i z_i e_i\right)\right\},
\]
where $w(0)=0$, $b\in B'$, $z_i\in\mathbb Z_{\ge0}$, $\sum z_ie_i\le N-b$, and $\sum z_i\le w(N)-2$. Indeed $b$ is the first sum of two old terms. The extras can then be used with the indicated multiplicities. The remainder is below $M$ and its minimum number of power-of-two terms is its binary weight. The pure-ladder upper bound makes the cap on $\sum z_i$ harmless. The bound $O(h^2\binom{d+w(N)-2}{d}\operatorname{poly}(d,D))$ reveals why the extra entries matter.

\paragraph{Numerical checks of the reduction and of suffix leaves.}
For two, three or four weights chosen nondecreasingly from $\{1,2,3,4,5\}$, all subset targets from zero to their sum gave $1{,}365$ checks of the dense construction ($913$ YES, $452$ NO); the same $1{,}365$ instances were unreachable with one fewer gate. Both a dense-tag and a separated-tag realization were tested. For the dense example $(2,4,7)$, $U=14$, the tags are $256,512,1024$ and $M=1045$. At $T=6$ use $256+14+2=272$, $512+14+4=530$, $1024+14=1038$: $1038+272=1310$, then $1310+530=1840$. At $T=5$ the subset sum is impossible. Nevertheless $1839$ has $\lambda=10$ and $w=8$, so its unrestricted binary chain has length $17$.

The suffix template implementation was compared with direct continuation search on $40$ separate histories: $4{,}476$ decisions with four and $5{,}740$ with five operations remaining. All agreed, and $7{,}053$ returned witnesses were replayed. The numbers of distinct cumulative signatures for $r=1,\ldots,6$ were $1,3,9,25,74,234$: a signature records the coefficient and whether the input is one old term or an all-old constrained pair. The following byte counts compare full NO trees generated with identical branching and other pruning, changing only the exact suffix horizon $H_j$. Each row has a separately checked witness of length $k+1$, so the proven optimum is $k+1$.
\begin{center}
\begin{tabular}{rrrrr}\hline
$N$ & $k$ & $H_3$ & $H_4$ & $H_5$\\\hline
63&7&231&159&123\\71&8&1073&309&159\\89&8&813&309&159\\
127&9&4584&1090&310\\135&8&350&232&160\\191&10&29803&6301&1121\\
255&9&638&352&232\\382&10&3332&1635&767\\383&11&111539&29717&6301\\\hline
\end{tabular}
\end{center}
The $27$ original files contain $26{,}843$ nodes in total. For $383$ the numbers of nodes were $12{,}994,\ 4{,}732,\ 1{,}024$; for $191$ they were $4{,}628,\ 1{,}024,\ 170$. These are uncompressed JSON byte lengths for selected cases; a shorter certificate can require more computation per leaf. The tree checker and generator share the exact suffix rule, which is why the separate comparison against literal suffix search is useful. The backward checker, by contrast, does not call a chain search inside its interval leaves. Neither experiment bounds worst-case proof sizes.

\section{Effective stabilization on an unrestricted dyadic orbit}\label{dsi:horizon}
Suppose $2^tn$ is the first point on the orbit where a shortest chain meets the budget $\lambda(n)+t+a$. If $t>0$, what would contradict that first occurrence? A chain for $2^{t-1}n$ with one fewer addition would meet the budget one shift earlier. The last operation need not be a doubling, so we look inside the chain. We record the runs that supply the two operands of each sum, together with their relative shift. This fixes the normalized sums. The endpoint inequalities below tell us how far their runs can move while both operands remain available.

\paragraph{Number of exceptional gates.}
Let $m=\lambda(n)$ and let a shortest chain for $N=2^t n$ have $r\le m+t+a$ additions, including $f$ steps other than immediate doublings. The estimate $a_r\le 2^{r-f}\varphi^f$ from S1.1, with $\varphi=(1+\sqrt5)/2$, gives $f\log_2(2/\varphi)\le r-\log_2N\le a$. Hence $f\le4a$, including when $r=0$. At every unmarked step choose the predecessor doubling, even if the value has other decompositions.

\paragraph{Coordinates for the runs.}
The \(f\) marked gates delimit \(f+1\) runs of lengths \(h_i\ge0\), \(0\le i\le f\). Write \(v_i\) for the value at the start of run \(i\). For some \(p_j,q_j<j\),
\[
 v_0=1,\quad v_j=2^{u_j}v_{p_j}+2^{z_j}v_{q_j},\qquad
 0\le u_j\le h_{p_j},\quad 0\le z_j\le h_{q_j},\qquad
 N=2^{h_f}v_f .
\]
All gates have a path to the target: otherwise removing unused computations would shorten a shortest chain. Distinguish the first parent at each sum and define
\[
 \rho_0=0,\quad \rho_j=\rho_{p_j}+u_j,\quad E_i=\rho_i+h_i,\quad
 d_j=\rho_{q_j}+z_j-\rho_j,\quad c_i=2^{-\rho_i}v_i .
\]
Then \(c_0=1,\ c_j=c_{p_j}+2^{d_j}c_{q_j}\) and \(N=2^{E_f}c_f\). For the fixed ordered parents and offsets \(d_j\), an integer endpoint assignment is feasible if
\begin{equation}
 \rho_0=0,\quad 0\le\rho_i\le E_i,\qquad
 \rho_{p_j}\le\rho_j\le E_{p_j},\quad
 \rho_{q_j}\le\rho_j+d_j\le E_{q_j}.             \label{dsi:endpoints}
\end{equation}
Indeed, reconstruct the two shift counts as \(\rho_j-\rho_{p_j}\) and \(\rho_j+d_j-\rho_{q_j}\). Both are integers in their parent runs. Induction on \(j\) gives a program on positive integers with output \(2^{E_f}c_f\) and cost \(f+\sum_i(E_i-\rho_i)\). To make a conventional increasing chain, keep only ancestors of the output, merge equal values, then sort distinct values in increasing order. Each noninitial value has a decomposition into two strictly smaller retained values. This never increases length.

\paragraph{A bound on offsets from binary carries.}
Unfold the \(f\) sums of the run diagram into a tree. There are at most \(M=2^f\) leaves, each contributing a nonnegative integer power of two to \(N\). Merge two equal powers to obtain one power with exponent one greater; continue until the ordinary binary expansion remains. At most \(M-1\) merges occur. Every final bit has an exponent in \([t,t+m]\), so every original exponent lies in \([t-(M-1),t+m]\). Following only distinguished edges gives leaf exponent \(E_f\). For any fixed gate \(j\) choose a path from it to the output; follow that same suffix after using the all-distinguished path to either of its parents. The two full path exponents differ by \(d_j\). In particular
\[
   |d_j|\le C:=m+M-1,\qquad E_f-t\ge-(M-1).                \tag{H1}
\]
The choice of a path upwards is legitimate because each gate is used. The nonnegativity assertion concerns the original leaf exponents, not the normalized offsets, which can be negative.

\paragraph{Gap lemma.}
Let \(t\ge1\) and suppose two consecutive endpoint coordinates in \([0,E_f]\) are \(L<B\), with \(B-L>C\). Decrease every endpoint at least \(B\) by one, including those beyond \(E_f\). The anchor remains zero. Every constraint in~\eqref{dsi:endpoints} has the form \(x-y\le c\) with \(|c|\le C\), using the anchor for the lower bound zero. The only adverse change is \(x\le L\), \(y\ge B\). In that case \(x-(y-1)\le L-B+1\le-C\le c\). Thus all operand constraints persist.

No run length increases. Along the distinguished path to the output, consecutive intervals meet, their union contains \([0,E_f]\), and no endpoint lies strictly between \(L\) and \(B\). Hence at least one interval crosses the gap and loses a doubling. The coefficients \(c_i\) are unchanged, so the new output is \(2^{E_f-1}c_f=2^{t-1}n\). Sorting the resulting program gives a chain with at least one fewer addition. When \(E_f>(2f+1)C\), such a gap exists by counting the at most \(2f+2\) endpoints between 0 and \(E_f\). This also covers \(f=0\) and \(C=0\).

Here is how the endpoint bound becomes a stopping rule. At a positive first feasible shift there is no such gap, so $E_f\le(2f+1)C$. By (H1), $t\le E_f+M-1$. Substituting $M=2^f$, $C=m+2^f-1$ and finally $f\le4a$ gives $H(n,a)$ of the finite-horizon theorem in the main article. If the budget is ever met, it is met by this horizon, and appending doublings carries it to the horizon itself.

\paragraph{Refinements and a worked cut.}
If \(v=\nu_2(n)\), \(w(n)\) leaves survive carrying, so at most \(M-w(n)\) merges are needed. In (H1) one can take \(C=m-v+M-w(n)\) and \(E_f-t\ge v-(M-w(n))\). For a specified program, the actual \(\max_j|d_j|\) also suffices in the cut (zero for an empty set of offsets). A gap can be shortened by \(\min\{t,B-L-C\}\) units at once. For \(n=15\), take runs of lengths \((t+1,2,0)\) giving \(v_1=2^t+2^{t+1}\) and \(v_2=v_1+4v_1=2^t15\). With the first operand distinguished at each gate the offsets are 1 and 2. For \(t>2\), shortening the internal gap yields \(1,2,4,8,12,24,48,60\), a chain to \(2^2 15\) of length seven. This illustrates cutting through an internal run even though the final run has length zero.

\subsection{Qualitative decidability from Flammenkamp's polynomial description}\label{dsi:flammen-decision}
Altman suggested using Flammenkamp's classification to decide stability \cite[\S2.2]{altmanint}. We give the implication explicitly. For each integer $a\ge0$, Flammenkamp's theorem \cite{flammenkamp}, in the form of \cite[Theorem~4]{altmanint}, gives an effectively computable finite family of polynomials with nonnegative integer coefficients. The positive integers $x$ with $s(x)\le a$ are the values
\[
 x=2^{u_0} f(2^{u_1},\ldots,2^{u_r}),\qquad
 u_0,\ldots,u_r\in\mathbb Z_{\ge0},
\]
over this family. Fix $n\ge1$, and let $B$ be the set of positions of its binary ones. We decide whether $x=2^t n$ occurs for some $t\ge0$.

Expand each positive coefficient of $f$ in binary. Each resulting occurrence of a power of two has an exponent of the form $u_0+\alpha_1u_1+\cdots+\alpha_ru_r+j$, where $j$ is a bit position of the coefficient. The number of such occurrences is fixed by $f$. Enumerate all rooted binary forests with these labelled occurrences as leaves, two children per internal node, and roots in bijection with $B$; a single leaf may form a tree. For each forest introduce an integer exponent at each vertex, fix the leaf exponents to the displayed affine forms, and require both children of every internal vertex to have exponent one less than their parent's. The root indexed by $b\in B$ has exponent $t+b$. Together with $t,u_i\ge0$, these are linear integer constraints, whose feasibility is decidable. The list of forests is finite, including the possibility that it is empty.

A solution gives the target sum since each merger preserves its value. Conversely, if the evaluated polynomial gives $2^t n$, repeatedly merge any two equal powers. The summand count decreases at each merger. The distinct powers left at termination are the binary expansion of $2^t n$, and the recorded mergers form one of the enumerated forests. This proves the decision criterion in both directions.

Apply the test for $a=0,\ldots,w(n)-1$. The binary method makes the last budget feasible at $t=0$. Let $a_*$ be the least feasible budget. Since $s(2^t n)$ is integer-valued and nonincreasing in $t$, its eventual value is $a_*$. Consequently $\ell_{\rm st}(n)=\lambda(n)+a_*$. To compute $K(n)$, test $t=0,1,\ldots$ at the fixed budget $a_*$ until the first feasible shift, which exists by the first test. This also decides stability by checking feasibility at $t=0$ alone for this budget. The argument establishes qualitatively what follows from the polynomial description. The gap and two-path arguments give explicit upper bounds for the first feasible shift in terms of the original chain parameters.

\section{From small-step catalogues to exact profiles}\label{dsi:small}
For an even number $N$ in a class with at most four small steps, we test whether $N/2$ already belongs to such a class. This local test identifies a first entry because the excess never increases under successive doublings. We use Knuth's description through two small steps, the three-step classification originating with Flammenkamp, and Clift's published data \cite{knuth,flammenkamp,clift,cliftsmall,cliftfour}. Each row represents a family of integers. If $p=(p_0,\ldots,p_{w-1})$ lists the set-bit positions of $N$ in decreasing order, halving subtracts one from every coordinate. A point that thereby leaves one row may still satisfy another. S5.2 checks membership in the union of the rows at this boundary. The conclusion for all integers requires the source catalogues to cover all the stated classes.

\subsection{A complete three-step test and its five-bit consequences}
The binary method is exact at weight at most three; $s=2$ permits at most four bits and $s=3$ at most eight \cite{knuth,altman,clift}.

For four set bits at positions $A>B>C>D\ge0$, Knuth's classification
\cite{knuth,altman,cliftsmall} has the following useful test:
$\ell(n)=A+2$ if at least one of
\[\begin{aligned}
 &A-B-C+D\in\{0,1\},\\
 &B=A-3\ \text{ and }\ D=C-1,\\
 &(A,B,C,D)=(D+7,D+2,D+1,D)
\end{aligned}\]
holds, and $\ell(n)=A+3$ otherwise. These are classical criteria, not a
new classification. In particular they are invariant under shifting all
positions by the same amount. The test distinguishes $23$ (length $6$)
from $29$ (length $7$), both of binary length five and weight four.

The low-weight and two-step classes are invariant under shifts whenever all positions remain nonnegative. In the 150 three-step rows, every equality is invariant under \(p\mapsto p+t\mathbf1\). There are eight inequalities with nonzero slope; each has slope one. Modulo the equalities within its row, its left-hand side after normalization to a nonnegative expression is
\[
 \begin{array}{ll}
 p_{w-1}-1,&\text{rows }22,30,36,37,42,45,112,\\
 p_{w-2}-2,&\text{row }69 .
 \end{array}
\]
Indeed, subtract the displayed affine form from the stored inequality: the difference is zero or a signed stored equality. All other inequalities have slope zero. Since \(p_{w-1}\ge0\), \(p_{w-2}\ge1\), a row that is ever reached is reached by the first doubling. Together with invariance of the lower levels this proves \(\min(4,s(2^tn))=\min(4,s(2n))\) for \(t\ge1\). If \(s(2n)=4\), this equality supplies a lower bound four subsequently, and monotonicity supplies the upper bound; the case \(s(2n)<4\) is immediate.

For five bits, the binary upper bound is four and the two-step classes have at most four ones. Only row 22 among the 26 five-bit three-step rows can change under doubling. Its positions satisfy
\[
 (a,a-7,b,b-1,b-3),\qquad b\ge4.
\]
To enter at the first doubling, the original unshifted pattern must have \(b=3\) and strict order forces \(a\ge11\). Substituting \((a,a-7,3,2,0)\) into the 26 rows gives just row 12 at \(a=11\) and row 4 at \(a=12\); each already has excess three. After a shift of one the entire half-line \(a\ge11\) is covered by row 22. Hence the unstable five-bit inputs are \(n=2^a+2^{a-7}+13\) with \(a\ge13\). The binary upper bound for \(n\) matches the exclusion of all cheaper classes. For \(2n\) put \(D=2^{a-6}\), take all powers of two through \(2D\), then append
\[
 3D,\ 3D+1,\ 6D+1,\ 9D+2,\ 15D+3,\ 30D+6,\
 60D+12,\ 120D+24,\ 129D+26.
\]
The last term is \(2n\) and the cost is \(a+4\). Further doublings meet the stable lower bound.

For the congruence criterion, if all \(p_i\) are congruent modulo \(d\), every integer linear combination of positions with coefficient sum zero is zero modulo \(d\). Thus the constant term of any required equality must be divisible by \(d\). For twenty of the 26 five-bit rows the greatest common divisor of their equality constants is one; for row 21 it is two. This eliminates all 21 rows when \(d\ge3\). Rows 4 and 23--26 remain. Each tests a different four-element subset. For positions \(x_1>x_2>x_3>x_4\) in that subset, its test is \(x_1+x_4=x_2+x_3\), or allows the two pair sums to differ by one in a prescribed direction. Congruence excludes the latter discrepancy. Conversely the exact equality always suffices. No other pairing can yield equal sums for four strictly ordered positions. This proves the equal-pair criterion in both directions. In the example \(\{0,1,2,4,8\}\), distinct positive powers of two have pair sums with two set bits, which cannot equal a sum involving zero; sums of distinct positive pairs are unique. The assumption \(d\ge3\) matters: \(16661\), with positions \((14,8,4,2,0)\), has no equal pair sums but has excess three by row 21.

\subsection{The four-step input and independent certificate checks}
The reduced file used here is Clift's \texttt{4\_r.txt} linked from the four-step enumeration \cite{cliftfour}. It has 40,206 rows and SHA-256 digest (join the two lines)
\[
\begin{array}{l}
\texttt{ba8d6d42eadfb192c940f70dc21080f5}\\[-2pt]
\texttt{544bc62eaab95c422310ee18cc5741b0}.
\end{array}
\]
All row identifiers refer to this version and the reduced three-step list used in the code. The strategy of inspecting a boundary and covering its halves with other rows is due to Clift \cite{cliftdouble}. He reported the broader bound \(k\le\lambda(m)-6\) for a drop ending at \(s\le5\), conditional on the enumerations, and noted that the floating-point infeasibility checks in that argument call for exact replay. Here the constant eight is restricted to an entry into level four and the checks use exact arithmetic. The number of sets reported by Clift for another run on the doubling page \cite{cliftdouble} is not used as a row count for this file. In using the four-step file we include the lower levels explicitly where needed. The arithmetic restrictions on \(p\) are equalities \(E=0\), inequalities \(G\ge0\), and any required divisibility conditions. Parameters exposed as individual bit positions can be substituted directly. For the remaining parameter, when present, solve a nonzero pivot equation; keep its denominator as a divisibility test and recheck all original equations after substitution. Direct parameter identification without such a pivot misses rows 1916, 2918, 4806, 6591 and 6593. The smallest and largest positions of an even output \(N=2^km\), for odd \(m\) and \(k\ge1\), obey
\[
    p_{w-1}=k,\qquad p_0=\lambda(m)+k.
\]

For an affine form \(G(p)=\sum_i g_i p_i+b\), put \(u=\sum_i g_i\). Then \(G(p-\mathbf1)=G(p)-u\). A row containing \(p\) can lose a defining inequality on halving only at \(0\le G(p)<u\) for \(u>0\). In the compiled file equalities and divisibilities are invariant under the translation. Distinctness and nonnegativity persist under halving because \(p_{w-1}\ge1\). There are 2,617 positive-slope inequalities. An audit for a proposed forbidden region \(D\) is the following finite proof obligation for each such inequality in row \(R\):
\[
 \left.
 \begin{array}{l}
 p\in R\cap D,\quad p_0>\cdots>p_{w-1}\ge1,\\
 0\le G(p)<u,\quad p-\mathbf 1\notin R'\ \text{for every selected covering row }R'
 \end{array}
 \right\}\quad\text{has no integer solution}.                 \tag{C1}
\]
An empty cover proves the boundary empty. Otherwise, (C1) proves that its entire half lies in the union of the selected rows. To verify (C1), start with the source constraints and process each cover row in turn. Write a target row as a conjunction \(Q_1\land\cdots\land Q_h\). Branch on the first failed constraint: branch \(i\) imposes \(Q_1,\ldots,Q_{i-1},\neg Q_i\), splitting a failed equality into its two strict alternatives. For $d>0$, failure of $d\mid A$ is the disjunction $A=dh+j$ for $1\le j<d$ and $h\in\mathbb Z$ (empty when $d=1$). These branches cover precisely the complement of the row. Discard a branch only after deriving an integer contradiction. If every branch is discarded after the last row, the cover is exhaustive.

For \(D=\{p_0-2p_{w-1}\le7\}\), 2,583 queries in the compiled catalogue are infeasible without a cover. Of the other 34, 24 have weight seven and ten have weight eight. Each has a cover using at most five rows. The covering row indices are recorded in \path{s9/theory/proof7.json}. For example, the certificate for inequality 0 of row 342 lists row 228 as its sole cover. For $750$ the vector is $p=(9,7,6,5,3,2,1)$, with $p_0-2p_6=7$. Halving changes the value of $-p_0+p_2+p_5+1$ in row 342 from $0$ to $-1$. The four left-hand sides in row 228, normalized against zero, take the values $1,1,0,0$ at $375$. The half thus leaves the first row and remains in the catalogue. The checker verifies coverage of the entire stated boundary. The verifier solves the corresponding contradictions by Fourier--Motzkin elimination. With integer variables an inequality \(a\cdot x\le b\) can first be tightened by dividing the left-hand coefficients by their positive greatest common divisor \(d\) and replacing \(b/d\) by \(\lfloor b/d\rfloor\). Every resulting contradiction is valid; no general completeness assumption on this elimination procedure is needed, because for these particular obligations it terminates with contradictions. A second implementation uses quantifier-free linear integer arithmetic in bit positions (Z3). The bound test and the equality test each give 2,617 unsatisfiable queries. The bound is about \emph{entering level four}; levels below it are dealt with by the preceding subsection.

For \(D=\{p_0-2p_{w-1}=8\}\), 61 boundary cases require explicit covers. Exclude the two stated candidate lines, cover the remainders by half-rows using \path{s9/theory/frontier8.json}, then intersect each exceptional line with all half-rows. In particular the seven-bit candidate at \(t=k\) is
\[
 (2k+8,\,2k+5,\,2k,\,2k-2,\,k+6,\,k+1,\,k).
\]
Among ordered candidates the half is covered for \(9\le k\le12\) and \(14\le k\le17\); precisely \(k=13\) and \(k\ge18\) yield a first entry. On the eight-bit ray the half qualifies for the exceptional candidates \(k=5,6,7,8\); all \(k\ge9\) yield first entry. Here \(C_j(p)\) denotes membership in the union through level \(j\). The ray substitution is an integer calculation. Each bit position becomes \(a_i k+b_i\). Substituting in a row leaves an integer interval cut out by inequalities and equalities, together with allowed residue classes from its divisibilities. Intersect and take the union of these sets separately for the target and its half, at levels three and four. Their Boolean difference \(C_4(p)\land\neg C_3(p)\land\neg C_4(p-\mathbf1)\) gives the ranges above, including the unbounded tails. The program \path{s9/check.py} recomputes these intersections in both coordinate systems. Beyond the last finite interval endpoint, one complete period of the remaining congruences suffices. The two rays are disjoint by weight. Since entry at lower levels can occur only for \(k=1\), separate checks of odd \(m\) with \(\lambda(m)\le8\) give the sole exception \(K_4(191)=1\) to the general bound on \(K_4\).

\subsection{Constructive chains at and before delayed savings}
For the eight-bit extremizer \(m=83\cdot2^{k+2}+71\), \(k\ge9\), the surviving four-step row 3244 is
\[
 (a,b,c,-a+b+c+1,d,-a+c+d+1,e,-a+b+e+1),\qquad -a+d+e\ge-1 .
\]
On the family one has \((a,b,c,d,e)=(k+8+t,k+6+t,k+3+t,6+t,1+t)\); thus the inequality reduces to \(t\ge k\). The other 4,707 rows of that weight and the 32 three-step rows are excluded by the translation-invariant tests on \(k\ge9\). Before entry the chain \(1,2,4,8,16,32,64,66,67,71,83\) prepares 71 and 83 in ten steps. Another \(k+2\) doublings and addition of 71 give \(m\) in \(k+13\) steps. At entry the following construction is Clift's \cite{cliftdouble}. Put \(x=2^k\), build the powers of two through \(2x\), and form \(3x,3x+1,5x+1\). Make four doublings to obtain \(80x+16\), add \(3x+1\) obtaining \(83x+17\), double \(k+2\) times, and add \(3x\). The output is \(2^km\) in \(2k+12\) steps. For the seven-bit extremizer, the chain
\[
 1,2,4,8,16,32,64,66,67,69,136,272,544,1088,1157
\]
prepares \(67,1157\) in 14 steps. Then \(k-2\) doublings and addition of 67 give \(m=1157\cdot2^{k-2}+67\) at cost \(k+13\). At the threshold, four-step membership gives the upper bound \(2k+12\). On either extremal family the lower-level test prohibits any subsequent drop below four (or use the one-doubling three-step theorem, since \(k>1\)). Thus the pre-entry length is \(k+t+13\) and the post-entry length is \(k+t+12\), including the exceptional seven-bit case \(k=13\).

For the six-bit family \(A_k=2^{2k+9}+2^{k+5}+2^{k+2}+21\), \(k\ge6\), the positions shifted by \(t\) fit just row 157 after translation-invariant constraints; all 52 three-step rows of weight six fail. Row 157 has positions
\((a,b,b-3,c,c-2,-a+2b-1)\) and inequality \(-a+b+c\ge0\). Substituting \((a,b,c)=(2k+9+t,k+5+t,4+t)\) gives \(t\ge k\). The binary method gives excess five before entry. At entry let \(x=2^k\), \(D=k+4\), \(X=2^D=16x\); start with powers through \(X/2\), then construct
\[
 A=3X/4,\quad B=5X/4,\quad C=B+1,\quad E=C+A .
\]
Form \(F=2^D E+C\) by \(D\) doublings and one addition, then \(2^kF+B\) in \(k+1\) more steps. This is \(512x^3+36x^2+21x=2^k A_k\) in \(3k+13\) steps.

For the seven-bit family \(B_k=2^{2k+8}+7\cdot2^{k+2}+21\), \(k\ge6\), membership of the four-step rows at \(t=k-1\) fails and at \(t=k\) succeeds by row 1363. The equalities of all three-step rows exclude the family for all \(t\ge0\). Therefore monotonicity proves the desired threshold on the entire orbit once we have an initial excess-five chain. Write \(B_k=(2^{k+6}+7)2^{k+2}+21\): build all powers through \(2^{k+6}\), inserting \(5,7,21\) in order (\(5=4+1,\,7=5+2,\,21=16+5\)); use two additions and \(k+2\) doublings for the final expression. This costs \(2k+13\). At entry put \(x=2^k\), \(h=8x^2\); start with powers through \(h\), then construct \(y=h+x,\ l=y+1,\ r=h+l,\ j=l+r,\ v=2j,\ z=v+r\). The target \(2^{k+2}z+y\) costs \(3k+12\). In all these constructions, appending doublings extends the chain to later shifts.

\subsection{Weight bounds, their equality cases, and finite extrema}
\begin{table}[ht]
\centering\small
\caption{Four-step row inventory. A moving inequality is one with positive slope under common translation of the bit positions. The final two rows list finite sharp delays and an odd base attaining each, where applicable.}
\begin{tabular}{c|rrrrrr}
weight&5&6&7&8&9&10\\ \hline
rows&1&226&1703&4708&7225&7621\\
moving inequalities&0&13&174&439&606&543\\
\(\sup K_4\)&1&\(\infty\)&\(\infty\)&\(\infty\)&3&3\\
witness for finite maximum&\(2^{13}+2^6+13\)&---&---&---&134987&40591\\[3pt] \hline
weight&11&12&13&14&15&16\\ \hline
rows&6464&5090&2532&2415&1373&848\\
moving inequalities&435&286&71&43&5&2\\
\(\sup K_4\)&3&3&1&1&1&1\\
witness for finite maximum&12251&285627&47087&49087&4962735&2482095
\end{tabular}
\end{table}
At weights one through four every relevant low-step condition is translation-invariant, so \(K_4=0\). At five the preceding classification is exact. At weights nine through twelve put \(T=3\); at thirteen through sixteen put \(T=1\). To verify the universal upper bound, it suffices to show that any row with an attainable future point on a given orbit already contains its \(T\)-th translate. All invariant constraints hold by eventual membership. For the positive-slope inequalities, apart from twelve cases the required shifted inequality follows directly from a lower bound on a single position and the decreasing order of nonnegative positions. The other twelve (two of weight nine, six of weight thirteen, four of weight fourteen) use explicit nonnegative rational combinations of valid inequalities and equalities, checked after clearing denominators. The coefficients of all twelve combinations and the code for checking their signs and resultant coefficients are in \path{s9/aux/}. The eight numerical witnesses in the table attain the bound. Weights above sixteen have no row through four steps and hence constant truncated value five. This last assertion does not amount to full stability.

At binary weight six, the stronger assertion is about the \emph{first entry} \(k>0\) into \(s\le4\). The lower bound \(\lambda(m)\ge2k+9\) now corresponds to excluding \(p_0-3p_5\le8\). There are 13 changing inequalities, and the 13 instances of (C1) are all unsatisfiable here. One can also see explicitly why this is a short audit: in nine cases the endangered constraint fixes \(p_5\) to 1 or 2, leaving a bounded enumeration; rows 99, 105, 157 and 158 leave unbounded candidates. Ordering and the boundary equation give small families whose halves all belong to rows 2, 7 or 9. The three-step classes might first appear at \(k=1\) only, and the six-bit odd bases with \(\lambda(m)\le10\) are already covered at \(t=0\).

There is a unique way to realize equality: \(k\ge6\) and \(m=A_k\). To check exhaustiveness, replace the bound on \(p_0\) by \(p_0=3k+9\). For the nine fixed-height inequalities test all ordered positions at the prescribed height. The four variable cases reduce, in order, to three, two, one and one rays: in row 99 the two nonnegative gaps satisfy \(u+v=2\); in row 105 they satisfy \(2u+v=2\); the remaining two rays are prescribed by their six positions. Checking all 226 rows at the half covers the admissible heights of every ray except row 157. Its admissible positions are
\[
 (3k+9,\,2k+5,\,2k+2,\,k+4,\,k+2,\,k),\qquad k\ge3.
\]
The half is covered precisely for \(3\le k\le5\). The separate lower-level check at \(k=1\) completes the equality test. Hence \(A_k\) is also the least odd six-bit base with first entry at \(k\) for every \(k\ge6\): it is the unique base of minimum bit length. For the five remaining positive heights the minima are not monotone:
\[
\begin{array}{c|rrrrr}
k&1&2&3&4&5\\\hline
\text{least }m&49429&263207&2099505&526407&4198669\\
\lambda(m)&15&18&21&19&22\\
\text{number at this minimum }\lambda&1&1&1&1&2
\end{array}
\]
The computation tests 26,082 six-bit odd bases with top position between 11 and 22. Each orbit is intersected symbolically with both catalogues. The lower bound excludes all smaller bit lengths. The equality classification and the small minima can be reproduced with \path{s9/aux/check_aux.py}.

For completeness, the finite maxima in the dyadic bands below the onset of the universal extremal formula are:
\[
\begin{array}{c|rrrrrrrrrrr}
d&0\text{--}6&7&8&9&10&11&12&13&14&15&16\\\hline
\max K_4&0&1&0&1&1&1&2&3&3&3&5
\end{array}
\]
All 65,536 odd bases below \(2^{17}\) enter this computation. It does not enumerate shifts up to a cutoff: each catalogue row, after substituting \(p+t\mathbf1\), gives a system in the single nonnegative integer \(t\), whose admissible set is solved exactly. For the 2,048 odd bases up to 4095 exact upper chains supplement the lower tests to give the \emph{full} profile on every shift. The nonconstant ones are
\[
 191,\ 701,\ 743,\ 1111,\ 1389,\ 1479,\ 2103,\ 2215,\ 2375,\ 2681,\ 2951 .
\]
Each drops by one in excess at the first doubling only. A useful certificate for the opposite behaviour is \(m=1903\): its orbit never enters level four, while
\[
 1,2,4,8,16,32,64,128,256,512,544,548,549,1098,1647,1903
\]
has length 15. Hence \(\ell(2^t1903)=15+t\) for \emph{every} \(t\). The same matching of an excess-five chain with an empty four-step orbit certifies all 39 bases for which a three-step lower bound and an upper chain alone are insufficient.

\paragraph{Recomputing the clipped and full profiles.}
The program \path{s9/bands.py} computes first entry at levels at most three and at most four on each of the $65\,536$ orbits and recovers all 17 band maxima above for the clipped profile. For weights at most four it uses the low-step classification from S5.1. At higher weights entry at level two or below is impossible. If $m<2^{14}$, the initial length table gives $s(m)\le5$ for all 8,192 odd bases. Monotonicity preserves this bound under every doubling. For these bases the first-entry times and initial cost determine the full pair $(K(m),\ell_{\rm st}(m))$. The program reproduces every row in the last scalar census of S6.

\paragraph{Verification materials and limitations.}
The public package \path{ASI_Verification_Lite.zip} includes compiled source predicates, cover certificates and independent checkers. The sharp $d-8$ statement assumes completeness of the source classifications \cite{cliftsmall,cliftfour}. The files for checking the arithmetic under this assumption are in \path{s9/}. The command \texttt{python s9/check.py --fetch} retrieves both source tables in a temporary directory, verifies their SHA-256 digests, recompiles every row and tests the covers by exact elimination. The second test, \texttt{python s9/check.py --z3}, runs offline on the supplied predicates. It solves 2,617 obligations for the bound and 2,617 for equality. The file \path{s9/README.md} also documents the use of local copies and export of the SMT-LIB queries. For decision-complexity statements input is binary unless sparse support is specified explicitly. The explicit output table for targets $1,\ldots,X$ has $X$ entries.

\section{Effective scalar horizons and certificates of stabilization}\label{scnew:section}
Let $m=\lambda(n)$, $w=w(n)$ be the binary weight, $v=\nu_2(n)$ and $u=n/2^v$. Write $s(x)=\ell(x)-\lambda(x)$ and $b_n(t)=\ell(2^tn)-t$. Appending a doubling gives $b_n(t+1)\le b_n(t)$, and $b_n(t)\ge m$. Its minimum is $\ell_{\rm st}(n)$ and its first minimizer is $K(n)$. A \emph{first hit} of level $a'$ is the least $t\ge0$ with $s(2^tn)\le a'$. Every strict drop first hits a lower level. Thus a bound valid for every positive first hit at levels $a'\le a=s(n)-1$ bounds $K(n)$. A certified upper bound for $s(n)$ can replace $s(n)$ in this choice. If $s(n)=0$, then $K(n)=0$. A chain witnessing a first hit below is always chosen shortest, with operands specified and every vertex ancestral to the output.
\subsection{Sharp growth at a fixed number of steps other than immediate doublings}
\begin{lemma}\label{scnew:fib}
Let $J_0=0$, $J_1=1$ and $J_{i+2}=J_{i+1}+J_i$. The largest possible endpoint of a length-$r$ chain with $f$ steps other than doublings of the immediately preceding term is $2^{r-f-1}J_{f+3}$, whenever the pair $(r,f)$ is feasible.
\end{lemma}
\begin{proof}
Immediately before a maximal block of $b\ge1$ such steps the last two values are $x/2,x$, since the first step of a nontrivial chain is a doubling. Each subsequent step gives at most the sum of the two largest distinct available terms. This also bounds a doubling of an older term. Relative to $b$ doublings the block achieves at most $J_{b+3}/2^{b+1}$. The identity $2J_{b+c+3}-J_{b+3}J_{c+3}=J_bJ_c$ follows from the Fibonacci recurrence (or induction on $c$). It bounds the product of two such ratios by the ratio for a single block of length $b+c$. Apply it successively. Equality is attained by $r-f$ initial doublings and $f$ successive sums of the two largest terms. For $f=0$ this includes the trivial chain.
\end{proof}
For a first hit at any $a'\le a$, write $r=\ell(2^tn)\le m+t+a$. The lemma gives the finite bound
\begin{equation}\label{scnew:F}
 f\le F=\max\{i\ge0:n2^{i+1}\le 2^{m+a}J_{i+3}\}.
\end{equation}
The ratios $J_{i+3}/2^{i+1}$ decrease to zero. An empty set excludes the hit. The count $f$ can include doublings of older terms and differs from $s(2^tn)$.
\subsection{Disjoint paths, carries and the number of leaves}
Represent each immediate doubling as a unary gate, every other step as a binary gate. An input--output path in the unfolded graph contributes $2^e$, where $e$ counts its unary gates. If $L$ is the number of leaves counted with multiplicity, merging equal powers of two requires $L-w$ carries. Each original leaf reaches a final bit at position at least $t+v$, so every path has $e\ge t+v-(L-w)$.
The same doublings are subject to two counts. The binary expansion forces each path to contain many of them when $t$ is large. In the original graph only $r-f$ such gates were paid for in total. To compare these counts, we need two paths that share no doubling gate, even though they may share addition gates. The fact that this is the first affordable shift supplies those paths.

\begin{proposition}\label{scnew:pflow}
A positive first hit admits two paths disjoint on unary gates. If $p\ge2$ paths are disjoint on those gates, counted with multiplicity on paths containing no unary gate, then
\[
 (p-1)t\le m+a-f+p(L-w-v).
\]
\end{proposition}
\begin{proof}
A unary gate common to all paths could be replaced by its input. This halves every term of the expansion, producing $2^{t-1}n$ in at most $r-1$ steps. Deleting unused values, merging equal ones and ordering by value gives an increasing chain at level $a'$ one shift earlier. Now split every gate into an input and output node, assigning its internal arc capacity one at unary gates and two elsewhere. Assign capacity two to all wires and terminal connections. A cut of capacity one would consist of a common unary gate. Integral max-flow/min-cut \cite{fordflow} gives two paths. For $p$ paths the sum of their exponents is at most $r-f$ since each unary gate is counted at most once. Insert the preceding lower bound on each exponent and $r\le m+t+a$.
\end{proof}
For fixed $a$, both $f$ and $L\le2^f$ are bounded uniformly in $n$, by \eqref{scnew:F} and $n\ge2^m$. Three paths would give $t\le m/2+O_a(1)$. Thus in a family with $t=m+O(1)$ every sufficiently large first-hit graph has a minimum cut of two unary gates (use capacity three on the other arcs). Using the catalogue premise and the first-entry check on the eight-bit ray in S5.2, Clift's family $n_k=83\cdot2^{k+2}+71$, $k\ge9$, first hits $a=4$ at $t=k=m-8$ \cite{cliftdouble}. With this input the coefficient of $m$ in the two-path bound is best possible for fixed $a=4$.
\begin{lemma}\label{scnew:leaves}
For $u>1$ every positive first hit has $f\ge2$ and
\begin{equation}\label{scnew:M}
 L\le M(f)=\min\!\left\{3\cdot2^{f-2},\;2^{f-1}+2^{\min(f,\Omega(u))-1}\right\},
\end{equation}
where $\Omega$ counts prime factors with multiplicity.
\end{lemma}
\begin{proof}
Group consecutive immediate doublings into runs, including a possibly empty run after each binary gate. The run graph has source $0$ and one vertex $j=1,\ldots,f$ for each new run base. If both parents of every gate belong to a single run, the pruned graph is a path. Each gate reads the end of the preceding run (later unused doublings would contradict minimality) and a strictly earlier term of that run. Hence its multiplier is $2^h(1+2^b)$ with $b\ge1$. Together with any terminal doublings, the factors $2^h$ supply $t+v>0$ powers of two. Shorten the terminal run, or lower both operand positions at a gate with $h>0$ and halve the remaining program, saving one step. This contradicts a first hit. It also excludes $f\le1$, including chains of pure doublings. Otherwise the first gate $j$ reading distinct runs has at most $2^{j-1}+2^{j-2}$ leaves. Subsequent gates at most double this bound. This proves the first term in \eqref{scnew:M}.

A \emph{dominator} is a run vertex on every source--output path. In topological order consecutive dominators bound blocks of sizes $h_1,\ldots,h_b$, with $\sum h_i=f$. Every earlier vertex reaches a given dominator and every later vertex is reachable from it. All vertices are used, and an edge across a dominator would bypass it. Thus each block builds an integer multiple of its input, and after division by the input it is an optimal chain for the quotient $R$. A shorter replacement would shorten the full program. In particular $R$ cannot be a power of two. Its block contains a non-immediate step, whereas attaining $2^e$ in $e$ steps forces immediate doubling throughout. The product of these quotients times the final power of two is $2^tn$. Therefore $b\le\Omega(u)$.

Within a block of $h$ gates write $c_i$ for the local path count, $c_0=1$ and $c_i\le2^i$. We prove $c_h\le2^{h-1}+1$. The maximum through position $j=1$ is at most $2^{j-1}+1$. If the bound holds through $j<h$, a path bypasses $j$ by an arc $p\to k$ with $p<j<k$. At $k$ its contribution is at most $2^{j-1}$. The other parent's contribution is at most $2^{k-j-1}(2^{j-1}+1)$. The sum is at most $2^{k-1}+1$ since $(2^{j-1}-1)(2^{k-j-1}-1)\ge0$. All earlier counts are at most $2^{k-1}$, so induction can resume at $k$. At dominators the counts multiply. In $\prod_i(2^{h_i-1}+1)$ moving a unit from a smaller $h_i\ge2$ to a larger $h_j$ cannot decrease the product, since the successive factor ratios increase. Concentrating all excess in one block gives $2^{f-1}+2^{b-1}$.
\end{proof}
Put $c=m+a-f$. With $F\ge2$ and $w\le M(F)$, Proposition~\ref{scnew:pflow} gives the horizon $\max(0,m+a-F-2v+2(M(F)-w))$, because $2M(f)-f$ increases with $f\ge2$. Otherwise no positive first hit exists. For powers of two $K=0$. Every path contributes one power of two of exponent at most $t+c$, the available number of doublings. Carry equal powers only below $2^{t+c}$. Quotient and remainder upon division by $2^{t+c}$ show that the resulting list has $P_c(n)$ terms and minimum exponent $t+\min(v,c)$. When $c<0$ every surviving term equals the cap. At most $L-P_c(n)$ carries occurred in total, so every original path uses at least $t+\min(v,c)-(L-P_c(n))$ doublings. The two paths in Proposition~\ref{scnew:pflow} together use at most $t+c$ doublings. Hence
\begin{equation}\label{scnew:cap}
 P_c(n)\le L,\qquad t\le c-2\min(v,c)+2(L-P_c(n)),\qquad
 P_c(n)=
 \begin{cases}\lfloor n/2^c\rfloor+w(n\bmod 2^c),&c\ge0,\\n2^{-c},&c<0.\end{cases}
\end{equation}
Take the largest bound over admissible $f\le F$, replacing $L$ by $M(f)$ if only the factorization is known. An empty maximum or a negative horizon gives zero.
\paragraph{}
Recursion through proper factors.\\
An internal dominator splits an optimal program into optimal factors $2^{k_1}u_1$, $2^{k_2}u_2$ with odd $u_i>1$, $u_1u_2=u$ and $k_1+k_2=t+v$. If $k_i>K(u_i)$, lowering $k_i$ by one saves one operation, since the stable tail of that factor has unit slope. Concatenating the two factor programs after scaling gives the original level one shift earlier. Hence
\[
 t\le K(u_1)+K(u_2)-v.
\]
Take the maximum of these bounds over proper factorizations and of the horizon for graphs without an internal dominator (one block, $L\le2^{f-1}+1$). Known upper bounds on the factor thresholds suffice. If all proper odd factors are stable, every graph with an internal dominator is excluded at positive $t$.

\subsection{Topology and finite enumeration for a specified base}
In a chronological run graph let $B_i$ be the base and $h_i$ the number of doublings of run $i$, with $B_0=1$. Gate $j$ reads runs $p_j,q_j<j$. A positive first hit has $h_f=0$, since a terminal doubling is removable. For $0\le i<f$ put $d_i=1$ if $p_{i+1}=q_{i+1}=i$ and $d_i=0$ otherwise. In the former case the gate must use the current maximum and a term no larger than half that maximum, since the two other possibilities either double the maximum or fail to exceed it. Thus $h_i\ge d_i$. Set $Y_0=1$ and compute an upper bound $Y_j$ for $B_j$ by
\[
 Y_j=\begin{cases}
 3\,2^{h_{j-1}-1}Y_{j-1}, &p_j=q_j=j-1,\\
 2^{h_{p_j}}Y_{p_j}+2^{h_{q_j}}Y_{q_j},&\text{otherwise}.
 \end{cases}
\]
Each path contributes a positive term involving every $2^{h_i}$ at most once. Hence $Y_f(h)2^{-\sum h_i}$ is nonincreasing in each coordinate. Since $2^tn=B_f$ and $f+\sum h_i\le m+t+a$, a necessary topology test is
\begin{equation}\label{scnew:growth}
 n2^{f+\sum d_i}\le 2^{m+a}Y_f(d).
\end{equation}
The exact leaf counts follow from $L_0=1$, $L_j=L_{p_j}+L_{q_j}$. Enumerate all unordered parent pairs at each gate, discard unused runs and apply \eqref{scnew:growth}, the dominator test and \eqref{scnew:cap}. For an individual surviving topology with horizon $H_T$ enumerate all heights with $\sum h_i\le m+a+H_T-f$ and all operand positions in those runs. Retain increasing chains of the required cost and endpoint, removing realizations with a common unary gate. These are necessary tests for every positive first hit at every level $a'\le a$. The largest retained shift is an upper bound, and exact costs locate the first minimum within that horizon.

For $191$ we have $m=w=7$, $v=0$, $\ell(191)=11$, $a=3$, $F=7$. Without factorization, $L\le3\cdot 2^5=96$ gives $7+3-7+2(96-7)=181$. The prime odd part gives $M(7)=65$, hence the uncapped horizon is $7+3-7+2(65-7)=119$. For $f=7$ the cap is $2^3=8$ after removing the common factor $2^t$. Carrying up to that cap gives $191=23\cdot8+4+2+1$, so $P_3(191)=26$. At most $65-26=39$ carries remain available in the bound. The $f=7$ contribution to \eqref{scnew:cap} is $3+2\cdot39=81$, the largest of the admissible contributions. The table gives bounds after each filter. The topology counts in the fourth column include the feasibility test $L\ge P_c$.
\begin{center}\begin{tabular}{@{}rrrrrrr@{}}\toprule
$f$&$c$&$P_c$&Topologies&$\max L_f$&Bound with $M(f)$&Bound with $\max L_f$\\\midrule
4&6&8&4&9&8&8\\5&5&10&50&17&19&19\\6&4&15&63&23&40&20\\7&3&26&5&34&81&19\\\bottomrule
\end{tabular}\end{center}
No smaller $f$ qualifies. The resulting horizon is $20$. Of the concrete programs, four pass the final tests, two each at $f=5,6$, all at $t=1$. Since $\ell(382)=\ell(191)=11$, the normalized cost drops from $11$ to $10$ there. Thus $K(191)=1$. An example optimal chain is $1,2,4,5,9,14,23,46,92,184,198,382$. With $5=4+1$, $9=5+4$, $14=5+9$, $23=14+9$, $198=184+14$ and $382=198+184$, two paths are
\[
\begin{aligned}
 1&\to\mathbf{2}\to\mathbf{4}\to5\to14\to198\to382,\\
 1&\to5\to9\to23\to\mathbf{46}\to\mathbf{92}\to\mathbf{184}\to382 .
\end{aligned}
\]
The paths share the sums $5$ and $382$, and their doubling gates (bold) are disjoint.

As a check, a second enumeration groups \emph{all} gates with identical operands into doubling paths. These form disjoint paths because a value has at most one double in an increasing chain. If $h$ is the number of remaining sums, $h\le F$. With $W_0=1$, assign $W_j=W_p+W_q$ for distinct parent paths and $W_j=3W_p/2$ on the same path (the two operands there are distinct). Expanding the recurrence for the actual output and dividing each path by $2^D$, where $D$ is the total number of doublings, gives an upper bound $W_h$. A path encounters each parent path at most once, and at a same-path sum the smaller operand is at most half the larger. Thus $W_h\ge n2^{-(m+a-h)}$. Applying this filter, \eqref{scnew:cap} with $f$ replaced by $h$, and finite enumeration gives horizon $26$ before operand enumeration and again only $t=1$ afterwards. For $127$ and $223$ the chronological enumeration excludes every positive hit at $a=3$, proving $K=0$ although the multiplier certificates below fall one step short.
\subsection{The strongest multiplier certificate for the low-weight bound}
Define $B(x)=\ell(x)$ for $w(x)\le4$ by the low-weight classification in S5.1. For $5\le w(x)\le8$ put $B(x)=\lambda(x)+3$, and for $w(x)\ge9$ put $B(x)=\lambda(x)+4$. By the low-step inequalities and the shift-invariant four-bit criteria in S5.1, $B(x)\le\ell(x)$ and $B(2^t x)=B(x)+t$ \cite{knuth,altman}. The optimal certificate using this $B$ is
\[
 B_{\times}(n)=\max_{z\ge1}\bigl(B(zn)-\ell(z)\bigr)\le\ell_{\rm st}(n).
\]
Indeed, concatenate a program for $2^tn$ with a scaled program for $z$ and use $B(2^tzn)=t+B(zn)$. The mechanism of transfer from multiples also appears in Clift's examples \cite{cliftdouble}. When $w\le4$, $z=1$ already attains $\ell(n)$. For $w\ge5$ the optimum is $m+3$ or $m+4$. It is $m+4$ precisely if $w\ge9$ or some $j\ge1$ satisfies
\begin{equation}\label{scnew:mult}
 2^j(2^{m+1}-n)\le n,\qquad w((2^j+1)n)\ge9.
\end{equation}
To prove the criterion, write $h=\lambda(z)$. Then $\lambda(zn)\le m+h+1$ and $B(zn)\le\lambda(zn)+4$. With at least three set bits, $\ell(z)\ge h+2$, giving at most $m+3$. A power of two gives $B(n)$. A two-bit multiplier is a shifted $2^j+1$ with unshifted cost $j+1$, and the shifts cancel. It gives $m+4$ precisely when $\lambda((2^j+1)n)=m+j+1$ and the product has at least nine bits, yielding \eqref{scnew:mult}. Only $2^j\le n/(2^{m+1}-n)$ need be tried.

The first equality $b_n(t)=B_{\times}(n)$ certifies the exact pair $(\ell_{\rm st}(n),K(n))$. For $n=3885$, $w=8$ and $17n=66045$ has weight nine, so $B_{\times}(n)=16+4-5=15=\ell(n)$ and $K=0$. For $6909$ and $13789$ the respective initial costs are $(17,18,18)$ and $(18,19,19)$, meeting the corresponding bound at $K=2$. For $12251$ and $12623$ the four costs are $(18,19,20,20)$ in both cases, meeting $17$ after normalization at $K=3$. In the last case $3n=37869$ has weight ten and leading bit $15$, giving $15+4-2=17$. The 20-step witness for $8n=100984$ is
\[
 \begin{gathered}
 1,2,4,8,16,32,64,128,256,512,513,1025,1538,2563,5126,\\
 6664,11790,23580,47160,94320,100984.
 \end{gathered}
\]
\paragraph{Catalogue dependence of the finite results.}
For the $8192$ odd $n<2^{14}$ the two older multipliers of S2 certified $1805$ thresholds using the costs available through $100000$ \cite{oeis3313}. Optimizing the multiplier of the strengthened $B$ certifies $4715$ within the same table, and the witness above adds one. Two disjoint structural computations then settle a further $268$. The first takes unresolved bases whose proper nontrivial odd divisors are already proved stable, with $F\le6$ (or $F\le7$ for $n<512$), and rules out a positive hit for all $231$ selected bases. The second uses the certified thresholds of proper divisors and $F\le8$ below $1024$. It certifies $37$ further zeros. For $955$ the structural bound is one and the exact costs at shifts zero and one settle equality. In particular the $512$ odd bases below $1024$ give $509$ zeros and $K(191)=K(701)=K(743)=1$, independently of S5.
\begin{center}\begin{tabular}{@{}lrrrrr@{}}\toprule
Odd $n<16384$ & $K=0$ & $K=1$ & $K=2$ & $K=3$ & Unresolved\\\midrule
$B_{\times}$, original cost table & 4647&65&2&1&3477\\
$B_{\times}$, including $100984$ & 4647&65&2&2&3476\\
With both structural stages &4915&65&2&2&3208\\
With the band bounds of S5.4 &8119&69&2&2&0\\\bottomrule
\end{tabular}\end{center}
The third row comprises $4984$ thresholds independent of S5. The last row additionally uses the catalogue completeness premise and the finite band calculation of S5.4. For the dyadic bands $0\le d\le13$ that calculation bounds the stabilization of $\min(5,s)$ by three. Every initial $s(n)$ in this sample is at most five, so truncation preserves the entire orbit. The four costs for $n,2n,4n,8n$ suffice. Missing values through $8n=131064$ were obtained by enumerating increasing chains through length 20, trying every sum of two earlier terms and discarding branches incapable of reaching the interval by successive doublings. For candidates absent through length 20, an available 21-step construction proves optimality. The overlap with the original table contains 125 matching values. Under S5 exactly 73 odd bases save one step, and all others are stable. Shifting a base uses $K(2^jn)=\max(0,K(n)-j)$, since $b_{2^jn}(t)=j+b_n(t+j)$. Thus the conditional counts for all $1\le n\le16384$ are $16310,70,2,2$ at $K=0,1,2,3$. Both $K=2$ bases and both $K=3$ bases already have independent multiplier certificates.
\subsection{Removing layers and optimizing over the shift}
\paragraph{Maximum simultaneous division in one given program.}
Choose operands and write \emph{every} addition of two identical values as a unary doubling. Replace the value $x_i$ by $x_i/2^{k_i}$, with $k_0=0$. At sums of distinct operands require both parents and the output to have equal scales. At a doubling $i\to j$ require $0\le k_j-k_i\le1$. Increment one removes the gate, increment zero keeps it. Encode a constraint $k_j-k_i\le d$ by an arc $i\to j$ of length $d$. Thus sum connections have length zero both ways, and doubling connections have length one forward and zero backward. The maximum possible scale at every vertex simultaneously is
\[
 k_i=\operatorname{dist}(0,i).
\]
Indeed every path bounds its endpoint scale by its length. Conversely the distances satisfy all inequalities by the triangle inequality and are nonnegative integers. Induction along the original dependency graph verifies the new integer values. If the output distance is $k$, each original path to the output contains $k$ removed gates, so at least $k$ operations disappear. Capping every distance at $j$ attains any output shift $0\le j\le k$. This solves the specified gate-preserving transformation exactly. The difference constraints are related to retiming in synchronous circuits \cite{retiming}.

For $2=1+1$, $3=2+1$, $4=2+2$, $5=2+3$, $10=5+5$, $14=4+10$, no doubling is common to all paths. Bypassing the gates for $4$ and $10$ yields $2+5=7$. Of $47\,403$ pruned, operand-labelled diagrams of length at most eight and even output, $16\,641$ have a common doubling. The simultaneous method finds $718$ additional diagrams admitting division. None of those $718$ was shortest for its output in this sample.

\paragraph{Computing the stable pair without choosing a target shift.}
Fix $s_0\ge s(n)$. For each $f$ allowed by \eqref{scnew:F} at $a=s_0$, enumerate live chronological topologies with ordered parents $p_j,q_j$. Let $L_0'=2^f$. Reject $w>L_0'$ and form
\[
 V_f=\{b-i:b\in\operatorname{supp}_2(n),\;0\le i\le L_0'-w\}.
\]
Every leaf exponent minus $t$ belongs to $V_f$, because a leaf ends at a binary position $t+b$ after at most $L_0'-w$ carries. Take $p_j$ as the primary operand at gate $j$. Normalize the input and successive run bases as $2^{\rho_i}c_i$, with $c_0=1$, $\rho_0=0$ and each run ending at $2^{E_i}c_i$. The actual two operands have heights $\rho_j$ and $\rho_j+d_j$ in their respective runs. Along the path obtained by always following the primary operand backwards, the exponent of the distinguished leaf equals the height at which the run is read. The two primary paths to the operands of gate $j$ can be extended by the same suffix from that gate to the output. Their final leaf exponents differ by $d_j$. Thus it suffices to try the finite choices $d_j\in V_f-V_f$. Evaluate $c_j=c_{p_j}+2^{d_j}c_{q_j}$ by exact dyadic arithmetic. Discard choices unless $c_f/n=2^\delta$ for some integer $\delta$.

For each surviving system, minimize $(r-t,t)$ lexicographically subject to
\begin{equation}\label{scnew:lp}
\begin{gathered}
 \rho_0=0,\quad 0\le\rho_i\le E_i,\quad
 \rho_{p_j}\le\rho_j\le E_{p_j},\quad
 \rho_{q_j}\le\rho_j+d_j\le E_{q_j},\quad E_f\ge-\delta,\\
 r=f+\sum_{i=0}^f(E_i-\rho_i),\qquad t=E_f+\delta .
\end{gathered}
\end{equation}
An integral point yields a program of cost $r$ and output $2^{E_f}c_f=2^tn$. Sorting and merging equivalent values can only lower its cost, so $r-t\ge\ell(2^tn)-t\ge m$. The difference inequalities have an integral constraint matrix of network type. At a vertex, tight differences join each variable to a fixed zero along a spanning forest, giving integral coordinates. Every nonempty face has a vertex since the polyhedron lies in the nonnegative orthant. An unbounded descent in the first objective would yield a rational recession direction that, after clearing denominators, gives unbounded descent through integral feasible points, contradicting $r-t\ge m$. Minimize $t\ge0$ on the first optimal face. Both minima are attained at an integral vertex. The lexicographic minimum across all systems is precisely $(\ell_{\rm st}(n),K(n))$, since every feasible program gives an upper bound, while a shortest chain at the first stable shift occurs in the enumeration. This is a finite exact algorithm. The topology and relative-shift enumerations can be large.

The exact-arithmetic prototype was compared with direct integer enumeration on $336$ systems. For $n=3,5,7,9,14,15,28$ it certifies respectively $\ell_{\rm st}=2,3,4,4,5,5,6$, with $K=0$ in every case. On the single topology from the displayed chain for $382$ it gives $(10,1)$. The global conclusion follows separately from the horizon or from S2. The public verification package \path{ASI_Verification_Lite.zip} includes the scalar calculations in \path{extensions/flow} and \path{extensions/descent}. The conditional population uses the finite band calculation of S5.4. Sharpness of the leading coefficient uses the unbounded first-entry family from S5.2. Both deductions assume catalogue completeness.

\subsection{A portable certificate for one fixed run pattern}
\label{v13:network-section}

For a fixed run pattern, finding an optimal schedule can be separated
from checking its optimality. The following certificate uses
standard network duality. It does not replace the proof that the enumeration
covers all relevant topologies and relative shifts.

Fix one feasible run-pattern system \eqref{scnew:lp}, including its parents and relative shifts, and retain its notation $c_f=2^\delta n$, $t=E_f+\delta$ and $r=f+\sum_i(E_i-\rho_i)$.
An integer point specifies a program of raw cost $r$; pruning and merging may lower its decoded cost.

Let $x$ contain the coordinates $\rho_i,E_i$, using $\rho_0$ as the fixed zero vertex $o$, with $x_o=0$.
Represent each inequality $x_v-x_u\le a_e$ by an arc $e:u\to v$ of
integer length $a_e$. Let $b$ have coefficient $+1$ at
$E_0,\ldots,E_{f-1}$, coefficient $-1$ at $\rho_1,\ldots,\rho_f$,
and zero elsewhere. Then
\[
 r-t=f-\delta+b\cdot x,\qquad t=E_f+\delta.
\]

\begin{proposition}[A certificate for both optimization objectives]
\label{v13:network-cert}
Suppose the integer vector $x$ satisfies every constraint.
Let $z_e\ge0$ be integer arc weights with outgoing minus incoming
balance $b_v$ at every vertex. If
\[
 b\cdot x=-\sum_e a_ez_e,
\]
then $x$ minimizes $r-t$ within this fixed system.
Add the reverse arc $v\to u$ of length $-a_e$ for each
arc $e:u\to v$ with $z_e>0$.
If this enlarged graph has a path from $E_f$ to the anchor of length
$-x_{E_f}$, then $x$ also minimizes $t$ among all minimizers
of the first objective.
\end{proposition}
\begin{proof}
For any feasible $y$, summing the constraints with weights $z_e$
gives $-b\cdot y\le\sum_ea_ez_e$. Equality at $x$ proves optimality
of the first objective. At any other optimum the sum of the
nonnegative slacks, weighted by $z_e$, is zero.
Every positive-weight constraint is therefore tight, so its reverse
inequality holds on the entire optimal face.
Summing the inequalities along the stated path gives
$0-y_{E_f}\le-x_{E_f}$ and hence $y_{E_f}\ge x_{E_f}$.
The witness $x$ attains this bound.
\end{proof}

The certificate consists only of a feasible integer vector, a flow with the
required balance, and a path. Its checker need not solve an LP or replay
the search that found the witness. The network part checks each encoded
constraint, balance and path step. Replaying the actual integer program is
a separate semantic check.

For the run pattern of the chain
$1,2,4,5,9,14,23,46,92,184,198,382$, viewed as a shift
of base $191$, the certificate gives $(r-t,t)=(10,1)$.
It applies to that particular pattern. The global assertion
$(\ell_{\rm st}(191),K(191))=(10,1)$ still requires the preceding horizon
argument or a global lower bound.

\paragraph{Exact assignment instead of permutation enumeration.}
The dual flow can be found by shortest paths and an assignment problem
between the $f$ vertices of balance $+1$ and the $f$ vertices of balance
$-1$. The implementation uses a classical integer assignment method with potentials and augmenting paths, in the tradition of the Hungarian method \cite{kuhn}. It requires $O(f^3)$ arithmetic operations instead of enumerating $f!$ permutations; bit cost also depends on the integer sizes. The outer enumeration of topologies and relative shifts, and the complexity class of the main theorem, are unchanged.

The computational companion implements the replacement in
\path{extensions/flow/planowanie/free_shift.py}, with its integer assignment routine beside it.
The portable examples and their independent checker are in \path{extensions/v13};
\texttt{python}\ \path{extensions/v13/verify_all.py} reruns the added checks.
The assignment routine agrees with complete permutation enumeration in 71 test cases, including signed costs of scale $10^{200}$ and forbidden edges. Certificates for bases $3$ and $191$ pass the arithmetic checks; deliberate changes to a potential, flow and path are rejected. These checks concern the encoded fixed systems. Completeness of an outer enumeration remains a separate mathematical and computational obligation.

\section{Fixed target sets and prescribed symbol counts}\label{v13:fixed-targets}
\subsection{Distinct prescribed outputs and a shared carry budget}\label{v13:joint-section}

Let $L(T)$ denote the minimum number of additions from $1$ producing every member of a prescribed finite target set $T$. Let $T\subset\mathbb Z_{>0}$ have $h\ge2$ distinct elements. Write
\[
 n_{\max}=\max T,\qquad m=\lfloor\log_2n_{\max}\rfloor,\qquad
 b_T(t)=L(2^tT)-t .
\]
The function $b_T$ is integer-valued, nonincreasing, and bounded below by $m$. Indeed, one initial doubling followed by a scaled copy of a sequence for $T$ constructs $2T$, and any sequence reaching $2^tn_{\max}$ has length at least $m+t$. Define
\[
 L_{\rm st}(T)=\min_{t\ge0}b_T(t),\qquad
 K_T=\min\{t\ge0:b_T(t)=L_{\rm st}(T)\}.
\]
For a singleton these definitions reduce to $\ell_{\rm st}(n)$ and $K(n)$, whose effective bounds are proved in Section~\ref{dsi:horizon}.

For an integer $c$ define
\[
 P_c(n)=
 \begin{cases}
 \lfloor n/2^c\rfloor+w(n\bmod2^c),&c\ge0,\\
 n2^{-c},&c<0,
 \end{cases}
 \qquad Q_c(T)=\sum_{n\in T}P_c(n).
\]
Let $\alpha_p(c,T)$ be the sum of the $p$ smallest members of the multiset
$\{\min(\nu_2(n),c):n\in T\}$, for $1\le p\le h$.

\begin{theorem}
\label{v13:distinct-carry}
Suppose $t>0$ is the first shift with $b_T(t)\le m+a'$ for integers $0\le a'\le a$. Choose a shortest sequence at that shift, specify its operands, and represent immediate doublings as unary gates and all other additions as binary gates. Let $r$ be its length and $f$ its number of binary gates, and put $c=m+a-f$.

Expand each prescribed output $2^tn$ separately into powers of two, with $L_n$ leaves, and set $L=\sum_{n\in T}L_n$. There are two source--output paths that end at distinct prescribed outputs and share no unary gate. More generally, if there are $p\ge2$ such paths with pairwise distinct prescribed endpoints, then
\begin{equation}\label{v13:dt}
 (p-1)t
 \le c-\alpha_p(c,T)+L-Q_c(T).
\end{equation}
In particular
\begin{equation}\label{v13:dh}
 t\le c-\alpha_2(c,T)+h2^f-Q_c(T).
\end{equation}
\end{theorem}
\begin{proof}
Every gate of a shortest sequence is an ancestor of at least one prescribed output. A unary gate on every source--output path could be bypassed, dividing every target by two and saving at least one operation. Removing unused values, merging equal ones and sorting by value would produce the previous shift within the same excess budget, contradicting first attainment.

Split each gate into an entrance and exit vertex. Give the internal arc capacity one at unary gates and capacity two at other gates; give the original dependency wires capacity two. Join each prescribed output to a new sink by an arc of capacity one. A cut of capacity one cannot be a single terminal arc: the other distinct prescribed outputs remain reachable. It cannot be a unary arc either, by the preceding paragraph. Hence integral max-flow/min-cut \cite{fordflow} gives two unit paths, disjoint on unary gates, using distinct terminal arcs. A path is allowed to pass through another prescribed output before ending at its own terminal.

There are $r-f\le t+c$ unary gates in the whole program. Every expanded leaf is therefore a power of two of exponent at most $t+c$, and $t+c\ge0$. For a fixed output $2^tn$, carry equal powers only below that cap. The resulting capped binary representation has $P_c(n)$ terms, with smallest exponent $t+\min(\nu_2(n),c)$. Exactly $C_n=L_n-P_c(n)$ carries were made. A leaf participates in at most $C_n$ of them, so any path ending at this output contains at least
\[
 t+\min(\nu_2(n),c)-C_n
\]
unary gates.

For $p$ selected paths ending at distinct outputs, each $C_n$ is charged at most once. Their total number of unary gates is thus at least
$pt+\alpha_p(c,T)-\sum_{n\in T}C_n$ and at most $t+c$, proving~\eqref{v13:dt}. Each output expansion has at most $2^f$ leaves, which gives~\eqref{v13:dh}.
\end{proof}

The distinction between endpoints is essential to the improved charging. For one prescribed output, two paths may belong to the same carry tree, and the earlier estimate charges its carry budget twice. The theorem does not replace that scalar estimate when $h=1$. It also does not apply unchanged to an optimization in which the selected output values themselves are allowed to vary with the shift.

\begin{corollary}[An effective horizon for every prescribed target set]\label{v13:joint-horizon}
For an integer $a\ge0$, let $\mathcal F(n_{\max},a)$ be the finite set of integers $f\ge0$ satisfying
\[
 n_{\max}2^{f+1}\le2^{m+a}J_{f+3},
\]
where $J_0=0$, $J_1=1$ and $J_{j+2}=J_{j+1}+J_j$. For each $f$ put $c_f=m+a-f$. Retain only $f$ with $Q_{c_f}(T)\le h2^f$, and define
\[
 H_T(a)=
 \max\left(
 \{0\}\cup
 \left\{
 c_f-\alpha_2(c_f,T)+h2^f-Q_{c_f}(T):
 f\text{ retained}
 \right\}\right).
\]
Then some shift satisfies $b_T(t)\le m+a$ if and only if the shift $H_T(a)$ does. In particular, if an integer $A\ge L(T)$ is certified by an explicit sequence, then $K_T=0$ when $A=m$, and otherwise
\[
 K_T\le H_T(A-m-1),
\]
where $K_T$ is the first shift attaining the eventual minimum of $b_T$.
\end{corollary}
\begin{proof}
Lemma~\ref{scnew:fib}, applied to the largest prescribed output $2^tn_{\max}$, gives the displayed necessary condition on $f$. Its finite nature follows from $J_{f+3}/2^f\to0$. The same $H_T(a)$ bounds every positive first attainment of every level $m+a'$ with $0\le a'\le a$, because all estimates use only the upper budget $a$. Prepending further doublings and scaling the entire sequence reaches the test shift while preserving all prescribed outputs. For the last assertion, any positive stabilization time attains a value strictly below $b_T(0)=L(T)\le A$, hence a level at most $A-1$.
\end{proof}

\paragraph{Numerical effect of distinct endpoints.}
At the same $c$ and leaf budget, using the uniform loss $L-Q_c(T)$ on each of two paths would give
\[
 t\le c-2\min_{n\in T}\min(\nu_2(n),c)+2\bigl(h2^f-Q_c(T)\bigr).
\]
Taking the maximum over the same admissible $f$ gives the following comparison. The witnesses in Example~\ref{v13:joint-example-si} certify $A=11$ for the first two rows; appending a doubling to a chain for $191$ gives $A=12$ for the last row. In every row $a=A-m-1=3$.
\begin{center}
\begin{tabular}{@{}l r r r@{}}\toprule
$T$ & $A$ & uniform loss on both paths & distinct endpoints\\\midrule
$\{7,191\}$ & 11 & 457 & 230\\
$\{7,23,191\}$ & 11 & 703 & 353\\
$\{191,382\}$ & 12 & 434 & 218\\\bottomrule
\end{tabular}
\end{center}
These are sufficient bounds from the specified estimates, not exact stabilization indices or a comparison with all scalar refinements. Example~\ref{v13:joint-example-si} settles the first two exact thresholds at one by a separate lower bound and witness. The arithmetic, leaf counts and explicit unary-disjoint paths are checked in \path{extensions/v13/verify_distinct_terminal.py} of the companion.

\subsection{Fixed composition and the exact cost of block order}\label{v13:composition-section}
Let $\mathbf n=(n_1,\ldots,n_q)$ have positive integer entries and put $T=\{n_1,\ldots,n_q\}$. Here $q$ counts the distinct symbols, whereas $h=|T|$ counts the distinct numerical multiplicities. Repeated counts occur once in $T$ but retain their multiplicities in sums over symbols. In the reusable-string model of Section~6 of the main article, basic symbols are free and each binary concatenation costs one step. Transposition, proved there, gives
\begin{equation}\label{v13:transposition-si}
 M(\mathbf n)=q-1+L(T).
\end{equation}
The order of an optimal string may change with the scale; neither a fixed string nor a fixed plan is prescribed.

\begin{proposition}[Effective stabilization of a fixed composition]\label{v13:fixed-counts-si}
For every positive count vector $\mathbf n$, one has
\[
 M(2^t\mathbf n)=q-1+t+L_{\rm st}(T)\qquad(t\ge K_T).
\]
The threshold $K_T$ is computably bounded by Corollary~\ref{v13:joint-horizon} when $h\ge2$, and by the scalar horizon of Section~\ref{dsi:horizon} when $h=1$. It depends on the distinct numerical counts in $T$, not on how many symbols have each count.
\end{proposition}
\begin{proof}
One initial doubling followed by a scaled copy of a sequence for $2^tT$ gives $L(2^{t+1}T)\le L(2^tT)+1$. The integer $b_T(t)$ is bounded below by $\lambda(\max T)$, so its minimum is attained and persists. Substitute $L(2^tT)=t+L_{\rm st}(T)$ in~\eqref{v13:transposition-si}. For a singleton $T=\{n\}$ the threshold equals $K(n)$; all other cases have the effective bound just proved.
\end{proof}

\begin{lemma}[One block for each symbol]\label{v13:block-identity-si}
For pairwise distinct symbols $\sigma_1,\ldots,\sigma_q$ and $n_i\ge1$,
\begin{equation}\label{v13:block-cost-si}
 \operatorname{ASI}(\sigma_1^{n_1}\cdots\sigma_q^{n_q})
   =q-1+\sum_{i=1}^q\ell(n_i).
\end{equation}
\end{lemma}
\begin{proof}
Build each unary block by a shortest chain, then concatenate the $q$ blocks. This gives the upper bound.

For the lower bound, prune a plan to its output and unfold its uses into a binary tree. Every used string is a substring of the final block word. A gate $U\,|\,V$ introduces the same ordered pair of symbols at its cut in each occurrence. Each of the $q-1$ distinct boundary pairs $\sigma_i\sigma_{i+1}$ must appear at the cut of some gate: take the lowest common ancestor of the two boundary positions in the unfolded tree. Different boundary pairs require different gates. Those gates add no two positive counts in any single-symbol projection, since the supports of their two operands are disjoint.

For any other gate, its cut lies inside the unique run of one symbol, say $\sigma_i$. Only this symbol can have positive counts in both operands. Project all strings onto their $\sigma_i$ counts and omit zero contributions and unchanged values. This gives a program producing $n_i$ from $1$, with at most as many additions as gates whose cut lies inside the $\sigma_i$ run. Sorting distinct positive values gives an addition sequence, so at least $\ell(n_i)$ such gates are needed. These gate classes are disjoint for different $i$, and disjoint from the boundary gates. Their sum proves the lower bound.
\end{proof}

\begin{corollary}[Eventually exact linear penalty for block order]\label{v13:block-order-si}
Define
\[
 \Delta_{\mathbf n}(t)=
 \operatorname{ASI}(\sigma_1^{2^tn_1}\cdots\sigma_q^{2^tn_q})-M(2^t\mathbf n).
\]
For every $t\ge\max\{K_T,K(n_1),\ldots,K(n_q)\}$,
\begin{equation}\label{v13:block-tail-si}
 \Delta_{\mathbf n}(t)=(q-1)t+\sum_{i=1}^q\ell_{\rm st}(n_i)-L_{\rm st}(T).
\end{equation}
In particular, $\Delta_{\mathbf n}(t+1)-\Delta_{\mathbf n}(t)=q-1$ on this tail. Its onset has a computable upper bound for every positive count vector.
\end{corollary}
\begin{proof}
Subtract~\eqref{v13:transposition-si} from~\eqref{v13:block-cost-si} at scale $2^t$. At every shift this gives
\[
 \Delta_{\mathbf n}(t)=\sum_i\ell(2^tn_i)-L(2^tT).
\]
Past all stated thresholds, the summands are $t+\ell_{\rm st}(n_i)$ and the joint term is $t+L_{\rm st}(T)$. Substitution gives~\eqref{v13:block-tail-si}. The effective joint and scalar horizons bound all thresholds in the maximum. Their cost values can be determined by finite exact enumeration; no efficiency claim is required for this computability conclusion.
\end{proof}

For $q\ge2$ the penalty diverges at the stated exact slope for every fixed positive frequency vector. Both the block string and an order attaining $M(2^t\mathbf n)$ have the same empirical single-symbol entropy
$-\sum_i(n_i/N)\log_2(n_i/N)$, where $N=\sum_i n_i$, independent of $t$. This observation concerns single-symbol frequencies, not block entropy or the entropy rate of a stochastic process.

\begin{example}[Fixed and free compositions can stabilize at different shifts]\label{v13:joint-example-si}
The two eleven-step chains
\[
\begin{gathered}
 1,2,3,4,7,14,21,23,46,92,184,191,\\
 1,2,4,5,9,14,23,46,92,184,198,382
\end{gathered}
\]
contain, respectively, $\{7,191\}$ and $\{14,382\}$. The exact values $\ell(191)=\ell(382)=11$ \cite{knuth,cliftdouble} imply
$L(\{7,191\})=L(\{14,382\})=11$.
The classical low-step implications
$s(n)\le1\Rightarrow w(n)\le2$ and
$s(n)\le2\Rightarrow w(n)\le4$ \cite[Theorem~4.4]{altman}
give $\ell(2^t191)\ge10+t$, because $191$ has seven binary ones and $\lambda(191)=7$. For $t\ge1$, prepend $t-1$ doublings and scale every entry of the second chain by $2^{t-1}$. This produces both required targets in $10+t$ additions. Consequently
\[
 L_{\rm st}(\{7,191\})=10,\qquad K_{\{7,191\}}=1,
 \qquad M(7,191)=12,
\]
and $M(2^t7,2^t191)=11+t$ for $t\ge1$.
The chain $1,2,3,6,7$ and the same low-step facts give $\ell(2^t7)=4+t$. Thus $\ell_{\rm st}(7)=4$, $\ell_{\rm st}(191)=10$, and
\[
 \Delta_{(7,191)}(t)=t+4\qquad(t\ge1).
\]
The displayed chains also supply $T=\{7,23,191\}$ and $2T=\{14,46,382\}$. The same lower bound and scaling prove $K_T=1$ and $L_{\rm st}(T)=10$ for that three-element set.

To compare with free composition, let $F_2(N)$ minimize ASI over all length-$N$ strings using two symbols. For $N\ge2$, $F_2(N)=\ell(N)$: projecting a string plan onto its total length proves the lower bound; if the last step of a shortest chain is $N=u+v$, its preceding steps supply both $u,v$, and~\eqref{v13:transposition-si} gives the reverse bound. The nine-step chain
\[
 1,2,3,6,12,24,48,96,99,198
\]
and the four binary ones of $198$ give
\[
 F_2(2^t198)=\ell(2^t198)=9+t\qquad(t\ge0).
\]
Thus the free-composition sequence $F_2(2^t198)-t$ is constant from shift zero, whereas the prescribed-composition sequence stabilizes at shift one. In the notation of the main article, $K_2^{\rm str}(198)=0$.
\begin{center}
\begin{tabular}{@{}c l r r r r@{}}\toprule
$t$ & counts & $M$ & block ASI & $\Delta$ & $F_2$\\\midrule
0 & $(7,191)$ & 12 & 16 & 4 & 9\\
1 & $(14,382)$ & 12 & 17 & 5 & 10\\
2 & $(28,764)$ & 13 & 19 & 6 & 11\\\bottomrule
\end{tabular}
\end{center}
The companion checks the displayed constructions and explicit transposed words. It also independently excludes all increasing chains of at most ten steps for $191$ and $382$, trying every operand pair and pruning only branches that cannot reach the target by successive doublings. This finite check verifies the initial scalar values; the low-step inequalities and scaled constructions establish the infinite tails.
\end{example}

\subsection{Sharpness of the leafwise carry estimate}\label{v13:barrier-section}

\begin{lemma}[A barrier for bounds using only value, cap and leaf count]\label{v13:carry-barrier}
Let $N>0$ and $B\ge0$ be integers. Put
\[
 P=\lfloor N/2^B\rfloor+w(N\bmod2^B),\qquad
 e_0=\min(\nu_2(N),B).
\]
Every representation of $N$ as $L$ powers of two with nonnegative exponents at most $B$ has all exponents at least
\[
 e_0-(L-P).
\]
If $0\le C=L-P\le e_0$, this lower bound is attained. If $1\le C\le e_0$, it can be attained simultaneously by two leaves.
\end{lemma}
\begin{proof}
Carrying equal powers only below $2^B$ yields the canonical capped representation with $P$ terms and smallest exponent $e_0$. There are $L-P$ carries, so a leaf rises by at most that many positions.

Conversely start from the canonical representation and choose a term $2^{e_0}$. Split it into two halves; at each subsequent step split one of the two current smallest terms. After $C$ splits there are $P+C=L$ terms, the smallest exponent is $e_0-C$, and for $C\ge1$ two terms have that exponent. The condition $C\le e_0$ keeps all exponents nonnegative.
\end{proof}

For $0\le L-P\le e_0$, the bound cannot be improved using only $N$, the cap and $L$. When $L-P>e_0$, nonnegativity gives the sharper trivial bound zero. Distinct prescribed outputs supply additional information through separate carry trees; topology, ancestry and constraints on reuse are other possible sources. The lemma concerns representations by powers of two; it does not assert sharpness within the narrower class of optimal first-hit dependency graphs.